\documentclass{aa}
\usepackage[varg]{txfonts}
\usepackage{braket}	
\usepackage{amssymb}
\usepackage[T1]{fontenc}
\usepackage{wrapfig}
\usepackage{fixltx2e}
\usepackage{mathtools}
\usepackage{hyperref}
\usepackage{array}
\usepackage{esint}
\usepackage{booktabs}
\usepackage{color}
\definecolor{blueish}{rgb}{0.2, 0.2, 0.9}
\definecolor{greenish}{rgb}{0.2, 0.7, 0.2}

\begin{document}

\title{Impact of convection and resistivity on angular momentum transport in dwarf novae}
 \author{N. Scepi\inst{1},  G. Lesur\inst{1}, G. Dubus\inst{1} \and M. Flock\inst{2}}
 \authorrunning{N. Scepi et al.}
   \institute{Univ. Grenoble Alpes, CNRS, Institut de Plan\'etologie et d'Astrophysique de Grenoble (IPAG), F-38000, Grenoble, France
%  \email{nicolas.scepi@univ-grenoble-alpes.fr}
 \and
 Jet Propulsion Laboratory, California Institute of Technology, Pasadena, California 91109, USA}

\abstract{The eruptive cycles of dwarf novae are thought to be due to a thermal-viscous instability in the accretion disk surrounding the white dwarf. This model has long been known to imply enhanced angular momentum transport in the accretion disk during outburst.  This is measured by the stress to pressure ratio $\alpha$, with $\alpha\approx 0.1$ required in outburst compared to $\alpha\approx 0.01$ in quiescence. Such an enhancement in $\alpha$ has recently been observed in simulations of turbulent transport driven by the magneto-rotational instability (MRI) when convection is present, without requiring a net magnetic flux. We independently recover this result by carrying out PLUTO MHD simulations of vertically stratified, radiative, shearing boxes with the thermodynamics and opacities appropriate to dwarf novae. The results are robust against the choice of vertical boundary conditions. The thermal equilibrium solutions found by the simulations trace the well-known S-curve in the density-temperature plane that constitutes the core of the disk thermal-viscous instability model. We confirm that the high values of $\alpha\approx 0.1$ occur near the tip of the hot branch of the S-curve, where convection is active. However, we also present thermally-stable simulations at lower temperatures that have standard values of $\alpha\approx 0.03$ despite the presence of vigorous convection. We find no simple relationship between $\alpha$ and the strength of the convection, as measured by the ratio of convective to radiative flux. The cold branch is only very weakly ionized so, in the second part of this work, we studied the impact of non-ideal MHD effects on transport. Ohmic dissipation is the dominant effect in the conditions of quiescent dwarf novae. We include resistivity in the simulations and find that the MRI-driven transport is quenched ($\alpha\approx 0$)  below the critical density at which the magnetic Reynolds number $R_\mathrm{m}\leq 10^4$. This is problematic because the X-ray emission observed in quiescent systems requires ongoing accretion onto the white dwarf. We verify that these X-rays cannot self-sustain MRI-driven turbulence by photo-ionizing the disk and discuss possible solutions to the issue of accretion in quiescence.}


\maketitle

\section{Introduction}
A fundamental, yet challenging, issue in accretion theory is the transport of angular momentum. Indeed, for matter to accrete it needs to transfer its angular momentum outward. Historically, the transport of angular momentum has been parametrized by the dimensionless parameter $\alpha$, the ratio of the fluid stress (responsible for the transport) to the local thermal pressure \citep{Shakura}. Thus, this "$\alpha$-prescription" reduces the physics to setting a phenomenological value for $\alpha$ .

Dwarf novae (DNe)  provide the best observational constrains on $\alpha$ \citep{king2007}. DNe are binary systems where matter is transferred by Roche lobe overflow from a solar-type star to a white dwarf. Their lightcurves show periodic outbursts during which the luminosity typically rises by several magnitudes \citep{Warner}. According to the disk instability model (DIM, see  \citealt{lasota2001} for a review), these outbursts are caused by a thermal-viscous instability due to the steep  temperature dependence of the opacity when hydrogen ionizes around 7000\,K. During quiescence,  the disk fills up as mass is transferred from the companion, gradually heating up until this critical temperature is reached at some radius. This triggers the propagation of heat fronts through the disk, bringing it into a hot state with a high accretion rate. The disk then empties until the temperature falls below the critical temperature and a cooling front returns the disk to quiescence. The timescales involved are set by the ability of the disk to tranport angular momentum, providing an observational handle on $\alpha$. Outburst decay timescales imply $\alpha\sim 0.1$ \citep{kotko} whereas recurrence timescales imply $\alpha\sim 0.01$ in quiescence \citep{cannizzo1988,cannizzo2012}. Although it has long been known that the DIM requires transport to be more efficient in outburst than in quiescence \citep{1984AcA....34..161S}, the physical reason for this change in $\alpha$ has remained elusive.

It is now widely accepted that angular momentum transport in disks is due to turbulence driven by the development of the magneto-rotational instability (MRI, \citealt{balbus1991}). The properties of MRI-induced transport have been extensively studied over the past 25 years using local, shearing box simulations. Isothermal,  stratified local simulations with zero net magnetic flux show a universal value of $\alpha \sim0.03$ (\citealt{Hawley1996}, \citealt{simon2012}), comparable to the value required for quiescent DNe. Higher values of $\alpha\sim 0.1$ require an external net vertical magnetic flux \citep{Hawley1995}, whose origin in DNe is unclear. Isothermal simulations are convenient idealisations but do not account for  turbulent heating or radiative losses, and thus cannot be used to investigate the thermal equilibrium states of DNe. \cite{latter2012} carried out more realistic simulations by using an analytic approximate local cooling law in a non-stratified shearing box, and showed numerically that the disk is indeed thermally unstable in the conditions of DNe. Their zero net flux simulations give $\alpha \sim0.01$ in both cold and hot states. \cite{Hirose2014} performed the first simulations including radiative transfer, vertical stratification and the realistic thermodynamics appropriate to DNe. They found that convection increases  $\alpha$ to 0.1 in the hot state near the critical temperature, in the absence of net magnetic flux, providing a tantalizing solution to the change in $\alpha$ in DNe \citep{2016MNRAS.462.3710C}.

Another explanation for the difference in transport efficiency between hot and cold states was proposed by \cite{Gammie1998}. In the quiescent state of DNe, the plasma is expected to be largely neutral and thus the magnetic field decouples from the disk. The conductivity drops as electrons become scarce and increasingly collide with neutrals. With the electron fraction a strong function of the temperature, they pointed out that the MRI may not be able to grow or, at least, sustain fully developed turbulence in a quiescent DNe disk.

In light of these results, we have carried out numerical simulations (\S2) to assess the impact of convection (\S3) and resistivity (\S4) on the transport of angular momentum in DNe in the hot, outburst and cold, quiescent states (respectively). In particular, we present the first local MHD shearing box simulations with realistic thermodynamics and radiative transfer that include Ohmic diffusion and conclude on the role that X-rays from the white dwarf boundary layer may play in the ionization balance of quiescent disks.

\section{Methods}
We adopt the local, shearing-box approximation \citep{Hawley1995}, to simulate a vertically-stratified patch of accretion disk situated at a distance  $R_0 = 1.315\times10^{10}\:\mathrm{cm}$ from a $0.6\,M_\odot$ white dwarf, giving an angular velocity $\Omega(R_0)=5.931\times10^{-3}\:\mathrm{s^{-1}}$. These values are identical to those chosen by \citet{Hirose2014} to facilitate comparisons. The simulations include radiative transport in the flux diffusion approximation  and  thermodynamic quantities appropriate to the temperature and density regime sampled by DNe.

\subsection{Basic equations}
Curvature terms are not taken into account in the shearing box approximation and the differential Keplerian velocity is modelled as a linearized shear flow $v_y^0=-(3/2)\Omega x$, where the $x$, $y$ and $z$ directions correspond to the radial, azimuthal and vertical directions respectively. The set of equations in the co-rotating frame is:
\begin{gather}
\frac{\partial\rho}{\partial t}+\pmb{\nabla}\cdot(\rho\pmb{v}) = 0  \\
\begin{split}
\rho\frac{\partial\pmb{v}}{\partial t}+(\rho\pmb{v}\cdot\pmb{\nabla})\pmb{v} = -\pmb{\nabla}\left(P+\frac{B^2}{8\pi}\right)+\left(\frac{\textbf{B}}{4\pi}\cdot\pmb{\nabla}\right)\textbf{B} +\rho\big(-2\Omega\pmb{\hat{z}}\times\pmb{v}\\ 
+ 3\Omega^2x\pmb{\hat{x}}-\Omega^2z\pmb{\hat{z}}\big)
\end{split}\\
\frac{\partial E}{\partial t}+\pmb{\nabla}\cdot[(E+P_t)\pmb{v}-(\pmb{v}\cdot\textbf{B})\textbf{B}] = -\rho\pmb{v}\cdot\pmb{\nabla}\Phi-\kappa_P\rho c(a_RT^4-E_R) \\
\frac{\partial\textbf{B}}{\partial t} = \pmb{\nabla}\times(\pmb{v}\times\textbf{B}-\frac{4\pi}{c}\eta\textbf{J}) 
\end{gather}
The last three terms of Eq.~2 represent, respectively, the Coriolis force, the tidal force and the vertical component of the gravitational force; $\pmb{\hat{x}}$ and $\pmb{\hat{z}}$ are the units vectors in the $x$ and $z$ directions.%\geo{Il faut definir $\Phi$}

Radiative transfer is treated separately from the MHD step using an implicit time-stepping following \cite{Flock2013}. In this step, we solve the coupled matter-radiation equations in the flux-limited diffusion approximation 
\begin{gather}
\frac{\partial E_R}{\partial t}-\pmb{\nabla}\frac{c\lambda(R)}{\kappa_R\rho}\pmb{\nabla}E_R = \kappa_P\rho c(a_RT^4-E_R) \\
\frac{\partial \epsilon}{\partial t} = -\kappa_P\rho c(a_RT^4-E_R) 
\end{gather}
where $\rho$ is the density, $\pmb{v}$ the fluid velocity vector, $P$ the thermal pressure, $\textbf{B}$ the magnetic field vector, $\Phi = (-3x^2+z^2)\times\Omega^2(\mathrm{R_0})/2$ is the gravitational potential in the co-rotating frame, $\eta$ the Ohmic resistivity, $\textbf{J}=(c/4\pi)\pmb{\nabla}\times\textbf{B}$ the current density vector, $E=\rho\epsilon+0.5\rho\pmb{v}^2+\textbf{B}^2/8\pi$ the total energy, $\epsilon$ the internal energy, $P_t=P+\textbf{B}^2/8\pi$ the thermal pressure plus the magnetic pressure, $c$ the speed of light, $a_R=(4\sigma/c)$ the radiation constant with $\sigma$ the Stefan-Boltzmann constant, $T$ the temperature, $E_R$ the radiation density energy, $\kappa_P$ the Planck opacity and $\kappa_R$ the Rosseland opacity. The radiative energy flux, in the flux diffusion approximation, is $\textbf{F}_\mathrm{rad}= (c\lambda(R)/\kappa_R(T)\rho)\pmb{\nabla}E_R$. The flux limiter is defined as $\lambda(R)\equiv(2+R)/(6+3R+R^2)$ with $R\equiv |\nabla E|/(\kappa_R\rho E)$ \citep{turner2001}. We do not take into account radiation pressure as it negligible for the temperatures reached by DNe; \citet{Hirose2014} found that it contributes at most 6\% of the gas+radiation pressure. 

To close our set of equations we use the following equation of state (EOS) and internal energy function: 
\begin{gather}
P = \frac{\rho}{\mu(\rho,T)}k_BT \\
\epsilon = \epsilon(\rho,T)
\end{gather}
where $\mu$ is the mean molecular weight. To compute $\mu$ and $\epsilon$, we use pre-calculated tables and interpolate linearly between values of the tables. Tables are computed from the Saha equations assuming ionization equilibrium for the solar composition of \cite{grevesse1998} with an hydrogen abundance $X=0.7$ and a metallicity $Z=0.02$. The adiabatic index $\Gamma$, the entropy and the thermal capacity $C_v$ are similarly computed. 

We use the opacity tables of \cite{Ferguson2005}, which cover the low temperature region from $2.7 < \log(T) < 4.5$,  and OPAL \citep{Iglesias1996}, which cover the high temperature region from $3.75 < \log(T) < 8.7$. We use a linear interpolation to connect the two  and extend the resulting table, where necessary, using a zero-gradient extrapolation. Our tables of opacities and thermodynamic quantities are fully consistent with those used by \citet{Hirose2014}.

\subsection{Boundary conditions}
We use shear-periodic conditions in the $x$-direction, periodic conditions in the $y$-direction, and either periodic or modified outflow conditions for the $z$-direction to test their influence on the results (\citealt{Hirose2014} used only outflow conditions). Outflow conditions usually assume a zero-gradient extrapolation to the ghost cell if material is going outside of the box and prevent matter from outside to come into the simulation box. Following \citet{Brandenburg1995}, we also impose the magnetic field to point in the $z$-direction at the interface with the ghost cells. This circumvents a spurious increase in the magnetic field intensity that we observed, but did not further investigate, when using pure outflow conditions.

The mass in the shearing box may decrease due to the outflow boundaries. To avoid this, we normalize the total mass to the initial mass at each time step by multiplying the density by a corrective factor. For the run 442O, the mass flux is of the order of $\sim0.1 \mathrm{g\:cm^{-2}}$ per orbital period. The time-averaged advective flux at the vertical boundary $[\braket{\epsilon v_z}](\pm z_{max})$ remains negligible in outflow conditions (the averaging procedure is described in \S\ref{sec:diag}).

\subsection{Numerical method}
We solve the MHD equations on a 3D Cartesian grid with the conservative, Godunov-type code PLUTO  \citep{mignone2009}. We choose a second order Runge-Kutta time integration method. Constrained transport ensures that $\nabla\cdot\textbf{B}=0$. The Riemann solver is HLLD except where the pressure difference between two adjacent cell exceeds five time the local pressure, in which case we allow for shock flattening by using the more diffusive solver HLL. To solve the radiative transfer equations, we follow the same implicit scheme as \cite{Flock2013} except that we use the bi-conjugate gradient  solver KSPIBCGS \citep{Yang:2002wg} as implemented in PETSC \citep{petsc}, which we find to provide stabler, faster convergence than BiCGSTAB for our application.

In order to avoid very small time steps, we use floors of $10^{-6}$ and $5\times 10^{-2}$ of the initial mid-plane values for the density and the temperature respectively.  We find that floors are activated over $\la 2\%$ of the box and, thus, have no significant impact on the thermodynamical equilibrium of the simulations.

\subsection{Initial conditions\label{sec:init}}

The initial flow is Keplerian with a weak zero net flux magnetic field of the form $B_z \propto \sin(2\pi x)$.  We usually start with an isothermal layer  and assume hydrostatic equilibrium to fix the initial vertical density profile. We let the MRI develop and then trigger radiative transfer after 48 local orbits (time is normalized by the quantity $1/\Omega_0$ thus one orbital period is equivalent to $2\pi$ in code units). This time is sufficient for the MRI to saturate and for the disk to reach a quasi-steady state. The isothermal temperature $T_{c0}$ is set to the mid-plane temperature found by using the vertical structure code of  \citet{hameury1998}  for given surface density $\Sigma_0$ and effective temperature $T_{\rm eff}$. This code solves the vertical structure equations assuming an $\alpha$-prescription for the angular momentum transport and associated heating rate, radiative transfer in the diffusion approximation, and convection described by mixing length theory with a mixing coefficient $\alpha_{\rm ml}=1.5$ (based on models of the Sun). We then activate radiative transfer and let the disk equilibrate and reach a quasi-steady state (if there is one). 

Some runs are initialized from another simulation by changing the surface density manually instead of starting from an isothermal layer. This has proven useful close to the hydrogen ionization regime, where the disk easily undergoes critical heating/cooling. In these cases, starting from an isothermal state can cause the disk to miss the equilibrium state. Starting from a nearby quasi steady-state allows for a smoother transition to capture the thermal equilibrium.

\subsection{Runs and diagnostics\label{sec:diag}}

\begin{table*}[h!]
\resizebox{\textwidth}{!}{%
\begin{tabular}{ c | c | c l c l c l c l c l c l c l c l c l c l c l c l c | c | c | c | c | c | c | c }
\toprule
 Run & Restart & $\Sigma_0$ & $T_{c0}$ & $H/R_0$  & $[T_\mathrm{mid}]\:\pm\sigma_\mathrm{T_\mathrm{mid}}$ & $T_\mathrm{mid_{max}}$ & $T_\mathrm{mid_{min}}$ & $[T_\mathrm{eff}]$ $\pm$ $\sigma_\mathrm{T_\mathrm{eff}}$ & $\alpha$ $\pm$ $\sigma_{\alpha}$ & $[\tau_\mathrm{tot}]$ & $n$ &  $\frac{L_x}{H}$ & $\frac{L_y}{H}$ & $\frac{L_z}{H}$ & $t_\mathrm{avg}$ \\
\midrule
\multicolumn{16}{c}{Hot branch} \\
\midrule
432P & - & 4000 & 231194 & 5.62E-02 & 216384 $\pm$ 7862 & 238779 & 205844 & 27729 $\pm$ 3249 & 0.066 $\pm$ 0.014 & 39294  & 21 & 1.5 & 6 & 12 & 16\\
430P & - & 2900 & 184762 & 5.03E-02 & 168546 $\pm$ 9001 & 188999 & 151736  & 22584 $\pm$ 2218 & 0.047 $\pm$ 0.014 & 34026  & 21 & 1.5 & 6 & 12 & 95\\
429P & - &1030 & 104765 & 3.78E-02 & 98395 $\pm$ 4544 & 110410 & 91551 &14793 $\pm$ 1617 & 0.042 $\pm$ 0.015 & 20450  & 20 & 1.5 & 6 & 12 & 95 \\
431P & - & 750 & 91431 & 3.54E-02 & 86437 $\pm$ 3314 & 95825 & 79883 & 13268 $\pm$ 1070 & 0.041 $\pm$ 0.010 & 18443 & 21 & 1.5 & 6 & 12 & 80\\
431O & - & 750 & 91431 & 3.54E-02 & 87500 $\pm$ 3097 & 96656 & 83558 & 13128 $\pm$ 899 & 0.042 $\pm$ 0.008 & 17520  & 21 & 1.5 & 6 & 12 & 95\\
439P & - & 540 & 79831 &  3.30E-02  & 72093 $\pm$ 3953 & 83749 & 67827 & 10718 $\pm$ 1296 & 0.031 $\pm$ 0.014 & 22519 & 20 & 1.5 & 6 & 12 & 95\\
439O & - & 540 & 79831 & 3.30E-02 & 74220 $\pm$ 3082 & 81313 & 68523 &10920 $\pm$ 969 & 0.033 $\pm$ 0.010 & 19578 & 21 & 1.5 & 6 & 12 & 95\\ 
468P & - & 386 & 70561 & 3.11E-02 & 66985 $\pm$ 1882 & 72396 & 62772 & 10493 $\pm$ 791 & 0.042 $\pm$ 0.009 & 16169 & 22 & 1.5 & 6 & 12 & 95\\
468O & - & 386 & 70561 & 3.11E-02 & 66784 $\pm$ 3186 & 76966 & 62428 & 10281 $\pm$ 1046 & 0.042 $\pm$ 0.014 & 16755 & 21 & 1.5 & 6 & 12 & 95\\
470P & - & 275 & 63720 & 2.95E-02 & 56203 $\pm$ 2773 & 62911 & 52180 & 8842 $\pm$ 1003 & 0.036 $\pm$ 0.013 & 18288 & 20 &1.5 & 6 & 12 & 48\\
437P & - & 174 & 54545 & 2.73E-02 & 44630 $\pm$ 4281 & 52848 & 36837 & 8117 $\pm$ 633 & 0.046 $\pm$ 0.017 & 15323 & 19 & 1.5 & 6 & 12 & 60\\
437O & - & 174 & 54545 & 2.73E-02 & 47414 $\pm$ 2788 & 52911 & 40886 & 8245 $\pm$ 499 & 0.052 $\pm$ 0.011 & 12375 & 20 & 1.5 & 6 & 12 & 95\\
441P & 437P & 134 & - & 2.73E-02 & 42882 $\pm$ 3757 & 51820 & 34140 & 8161 $\pm$ 633 & 0.070 $\pm$ 0.021 & 10344 & 19 & 1.5 & 6 & 12 & 95\\
441O & 437O & 134 & - & 2.73E-02 & 35605 $\pm$ 7841 & 51140 & 18170 & 7795 $\pm$ 511 & 0.061 $\pm$ 0.022 & 19797 & 17 & 1.5 & 6 & 12 & 95\\
446P & 437P & 127 & - & 2.73E-02 & 30515 $\pm$ 9137 & 46504 & 15721 & 7563 $\pm$ 421 & 0.066 $\pm$ 0.027 & 25335 & 16 & 1.5 & 6 & 12 & 92\\
446O & 437O & 127 & - & 2.73E-02 & 34881 $\pm$ 5947 & 42975 & 17104 & 7567 $\pm$ 330 & 0.061 $\pm$ 0.021 & 16285 & 17 & 1.5 & 6 & 12 & 95\\
442P & 437P & 113 & - & 2.73E-02 & 35004 $\pm$ 6857 & 45903 & 17805 & 7745 $\pm$ 419 & 0.079 $\pm$ 0.031 & 12796 & 17 & 1.5 & 6 & 12 & 83\\
442O & - & 113 & 34500 & 2.17E-02 & 32724 $\pm$ 4878 & 43105 & 20887 & 7292 $\pm$ 323 & 0.075 $\pm$ 0.022 & 15318 & 21 & 1.5 & 6 & 12 & 95\\
452P & 437P & 100 & - & 2.73E-02 & 28719 $\pm$ 9205 & 41063 & 14888 & 7302 $\pm$ 435 & 0.092 $\pm$ 0.024 & 18108 & 15 & 1.5 & 6 & 12 & 95\\
452O & 442O & 102 & - & 2.17E-02 & 33245 $\pm$ 5735 & 44029 & 19624 & 7390 $\pm$ 348 & 0.087 $\pm$ 0.031 & 11514 & 21 & 1.5 & 6 & 12 & 95\\
453P & 437P & 90 & - & 2.73E-02 & 35292 $\pm$ 3918 & 42119 & 23611 & 7332 $\pm$ 627 & 0.098 $\pm$ 0.039 & 7357 & 17 & 1.5 & 6 & 12 & 84 \\
453O & 442O & 90 & - & 2.17E-02 & R & R & R & R & R & R & R & 1.5 & 6 & 12 & 95\\
\midrule
\multicolumn{16}{c}{Middle branch} \\
\midrule
438P & - & 275 & 14000 & 1.38E-02 & R & R & R & R & R & R & R & 1.5 & 6 & 12 & 48\\
428O & - & 229 & 13000 & 1.33E-02 & R & R & R & R & R & R & R & 1.5 & 6 & 12 & 95\\
403O & - & 220 & 13000 & 1.33E-02 & 7893 $\pm$ 1261 & 9956 & 5362 & 4369 $\pm$ 162 & 0.027 $\pm$ 0.010 & 2050 & 33 & 0.75 & 3 & 6 & 191\\ 
402O & - & 210 & 13000 & 1.33E-02 & 8085 $\pm$ 1370 & 10515 & 5016 & 4423 $\pm$ 141 & 0.029 $\pm$ 0.011 & 2560 & 35 & 0.75 & 3 & 6 & 318\\ 
401O & - & 200 & 12000 & 1.28E-02 & 9431 $\pm$ 1224 & 11964 & 6348 & 4514 $\pm$ 187 & 0.036 $\pm$ 0.012 & 8458 & 38 & 0.75 & 3 & 6 & 350\\ 
404O & - & 150 & 9000 & 1.11E-02 & R & R & R & R & R & R & R & 0.75 & 3 & 6 & 95\\
\midrule
\multicolumn{16}{c}{Cold branch} \\
\midrule
435P & - & 191 & 2645 & 6.01E-03  & 3783 $\pm$ 103 & 3983 & 3559 & 3206 $\pm$ 153 & 0.036 $\pm$ 0.009 & 16 & 15 & 0.75 & 3 & 6 & 95\\
435O & - & 191 & 4000 & 7.39E-03 & 3785 $\pm$ 127 & 4055 & 3593 & 3334 $\pm$ 213 & 0.036 $\pm$ 0.011 & 15 & 42 & 0.75 & 3 & 6 & 95\\ 
462P & - &174 & 2785 & 6.17E-03 & 3102 $\pm$ 262 & 3471 & 2643 & 2855 $\pm$ 179 & 0.042 $\pm$ 0.014 & 8 & 45 & 0.75 & 3 & 6 & 95\\
465P & 462P & 116 & - & 6.17E-03 & 2095 $\pm$ 115 & 2346 & 1922 & 2332$\pm$ 115 & 0.035 $\pm$ 0.008 & 2 & 39 & 0.75 & 3 & 6 & 95 \\
434P & - & 93 & 1976 & 5.20E-03 & 1897 $\pm$ 49 & 2028 & 1814 & 2115 $\pm$ 109 & 0.037 $\pm$ 0.010 & 1 & 42 & 0.75 & 3 & 6 & 95\\
434O & - & 93 & 1976 & 5.20E-03 &1861 $\pm$ 33 & 1966 & 1814 & 2022 $\pm$ 108 & 0.034 $\pm$ 0.010 & 1 & 42 & 0.75 & 3 & 6 & 95\\
476P & - & 45 & 1828 &  4.99E-03 & 1715 $\pm$ 47 & 1840 & 1604 & 1733 $\pm$ 116 & 0.037 $\pm$ 0.011 & 4 & 41 & 0.75 & 3 & 6 & 56\\
\bottomrule
\end{tabular}}
\caption{Initial parameters and results for our ideal MHD simulations. $\Sigma_0$  is the initial surface density in g\,cm$^{-2}$, $T_{c0}$ the initial midplane  temperature in K, and $H/R_0$ where $H$ is the corresponding scale height. Brackets $[]$ are for quantities averaged over $t_\mathrm{avg}$ (given in local orbits) with $\sigma$ the associated standard deviation. $L_{x}/H$, $L_{y}/H$ and $L_{z}/H$ are the number of initial scale heights in the box in the $x$, $y$, $z$ directions. $n$ is the number of points per  time-averaged scale height. The second column indicates the simulation from which the run was initialized, otherwise we started from an isothermal layer at $T_{c0}$. P denotes a run with periodic vertical boundary condition; O denotes outflow conditions. R signals simulations with runaway heating or cooling. 
}
\label{tab}
\end{table*}

Table \ref{tab} lists the runs that we performed. We adopt the notation of  \cite{Hirose2014}  for labelling the runs, with the addition of a final O to indicate simulations with vertical outflow boundary conditions or a final P for a periodic vertical boundary. $\Sigma_0$ is the initial surface density and  $H\equiv c_s(T_{c0})/\Omega$ is the pressure scale-height (with $c_s$ the sound speed). The horizontal extent of the box is of $\pm$6$H$ for the hot branch and $\pm$3$H$ for the cold and middle branch. $L_x,\:L_y$ and $L_z$ follow the ratio 1:2:4 (see Table \ref{tab}). 

The horizontal average of a quantity $f$ is defined as :  
\begin{equation}
\braket{f}_{x,y}(z,t) \equiv \frac{\iint f(x,y,z,t)\mathrm{d}x\,\mathrm{d}y}{\iint \mathrm{d}x\,\mathrm{d}y}
\end{equation}
where the integration in $x$ and $y$ is done over the whole simulation box. We use $\braket{}$ brackets to denote spatial averages and $[ ]$ brackets to indicate an average  over a time $t_\mathrm{avg}$ performed once the simulation has reached a quasi-steady state. The averaging timescale $t_\mathrm{avg}$ is indicated for each run in the last column of Table \ref{tab}. We typically average over a hundred orbits except for somes cases where the variability occurs on long timescales (e.g. 401O, see \S\ref{sec:cycle}).  $[\Sigma]$, $[T_\mathrm{mid}]$, $[T_\mathrm{eff}]$, and $[\tau_\mathrm{tot}]$ are thus the time-averaged values of the surface density, midplane temperature,  effective temperature and total optical depth, respectively. $\sigma_{T_\mathrm{mid}}$, $\sigma_{T_\mathrm{eff}}$ and $\sigma_{\alpha}$ are the standard deviations of the fluctuations with time of these quantities.

We compute $\tau_\mathrm{tot}$, $\Sigma$, $T_\mathrm{eff}$ as
\begin{equation}
\tau_\mathrm{tot} = \int{\braket{\rho}_{x,y}\braket{\kappa_R}_{x,y}dz}
\end{equation}
\begin{equation}
\Sigma=\int{\braket{\rho}_{x,y}dz}
\end{equation}
and
\begin{equation}
T_\mathrm{eff}=\left(\frac{1}{\sigma_B}(F_\mathrm{rad\:z}^+-F_\mathrm{rad\:z}^-)\right)^{1/4}
\end{equation}
where $F_\mathrm{rad\:z}^{+/-}$ is the radiative flux in the vertical direction on the upper/lower boundary of our simulation box.
We also define $\braket{Q^-}_{x,y}$ the total cooling rate as
\begin{equation}
\braket{Q^-}_{x,y}\equiv \frac{d}{dz}\braket{F_\mathrm{rad\:z}}_{x,y}+\frac{d}{dz}\braket{\epsilon v_z}_{x,y}
\label{eq:fluxes}
\end{equation}
$v_z$ the vertical velocity and the quantity $\epsilon v_z$ represents the advective flux of internal energy $F_\mathrm{adv}$. The integration over $z$ extends over the full box.

We define $\alpha$, the ratio of stress to pressure, as
\begin{equation}
\alpha \equiv \frac{[\braket{W_{xy}}_{x,y,z}]}{[\braket{P}_{x,y,z}]}
\end{equation}
and the instantaneous stress to pressure ratio as $\tilde{\alpha}=\braket{W_{xy}}_{x,y,z}/\braket{P}_{x,y,z}$.
The turbulent stress $W_{xy}$ is $\rho (v_xv_y-v_{\mathrm{A}x}v_{\mathrm{A}y})$ where $v_\mathrm{A}$ is the Alfv\'en velocity.

\subsection{Numerical convergence}
We take 32 points in the $x$ direction, $128$ points in the $y$ direction and $256$ points in the $z$ direction in all our simulations. This number of points must be sufficient to resolve the most unstable wavelength of the MRI, which is of order $H$ (as the root mean square turbulent magnetic field is of the order of 0.1 - 0.2 for all simulations), while still allowing for a significant vertical extension. Hence, we use different values of $L_x$, $L_y$ and $L_z$ on the cold branch and on the hot branch. We note $n$ the number of points per final time-averaged scale height $h\equiv c_\mathrm{s}([T_\mathrm{mid}])/\Omega$. On the high $\Sigma$ part of the hot branch $n$ is of the order of 20 (Tab.\,\ref{tab}), enough to resolve the MRI. But at lower $\Sigma$  some of our simulations are restarted from previous runs with higher temperatures, decreasing $n$ to 15 in run 452P.  We checked those cases though additional runs with a more appropriate box size and resolution and found that the results are quantitatively the same. For instance, we find a difference of $\sim5\%$ in $T_\mathrm{mid}$ and $\sim2\%$ in $\alpha$ between the run 442O started with an isothermal layer (with $n=21$) or initialized from run 437O (with $n=13$). 

High Alfv\'enic velocities are allowed near the boundaries when the box height is large, about $12H$, increasing the numerical diffusion due to the HLLD solver and providing a numerical source of heating. On the hot branch, the contribution of this numerical coronal heating with respect to the total numerical heating is negligible:  we performed the simulation 437O with $L_x/H=1.5$, $L_y/H=6.0$ and $L_z/H=8.0$ and found a difference on $[T_\mathrm{mid}]$ of only 2$\%$ for the hot branch. The effect is much stronger on the middle and cold branch, heating being weaker, leading us to take smaller box sizes. Consequently, $n$ is approximately two times larger in the cold branch than on the hot branch. 

According to \citealt{ryan2017}, isothermal stratified simulations are not fully resolved since $\alpha$ decreases as $n^{-1/3}$ up to the highest achievable resolutions. Therefore, a caveat when comparing runs on the cold and hot branch is that the two-fold increase in resolution may lower the value of $\alpha$ by $\sim20\%$ in the cold branch simulations. We ran the simulation 434O with half the resolution on each direction and found $\alpha=0.019$. This difference is of the same order as the estimated error on $\alpha$. Hence we cannot conclude about consistency with \citealt{ryan2017}. A proper convergence study would be useful to know if this result holds for our simulations including realistic thermodynamics and radiative transfer.

Finally, \citet{simon2012} showed that the box size $L_x$ and $L_y$ must be $\geq 2H$ to have converged diagnostics of the MRI.  Again, these results were obtained using an isothermal equation of state and we do not know for sure if they hold here.  According to this criterion, our box size is appropriate on the hot branch but may be too small in $x$ on the cold branch (Tab.\,\ref{tab}). In particular, convective eddies in our simulations could be limited by small box sizes. We doubled the size in $x$ and $y$ of the run 442O to check for this effect and found no notable difference.

\section{Ideal MHD runs}

\subsection{Thermal equilibrium curve}

\begin{figure*}[h!]
\begin{center}
\includegraphics[width=200mm,height=100mm]{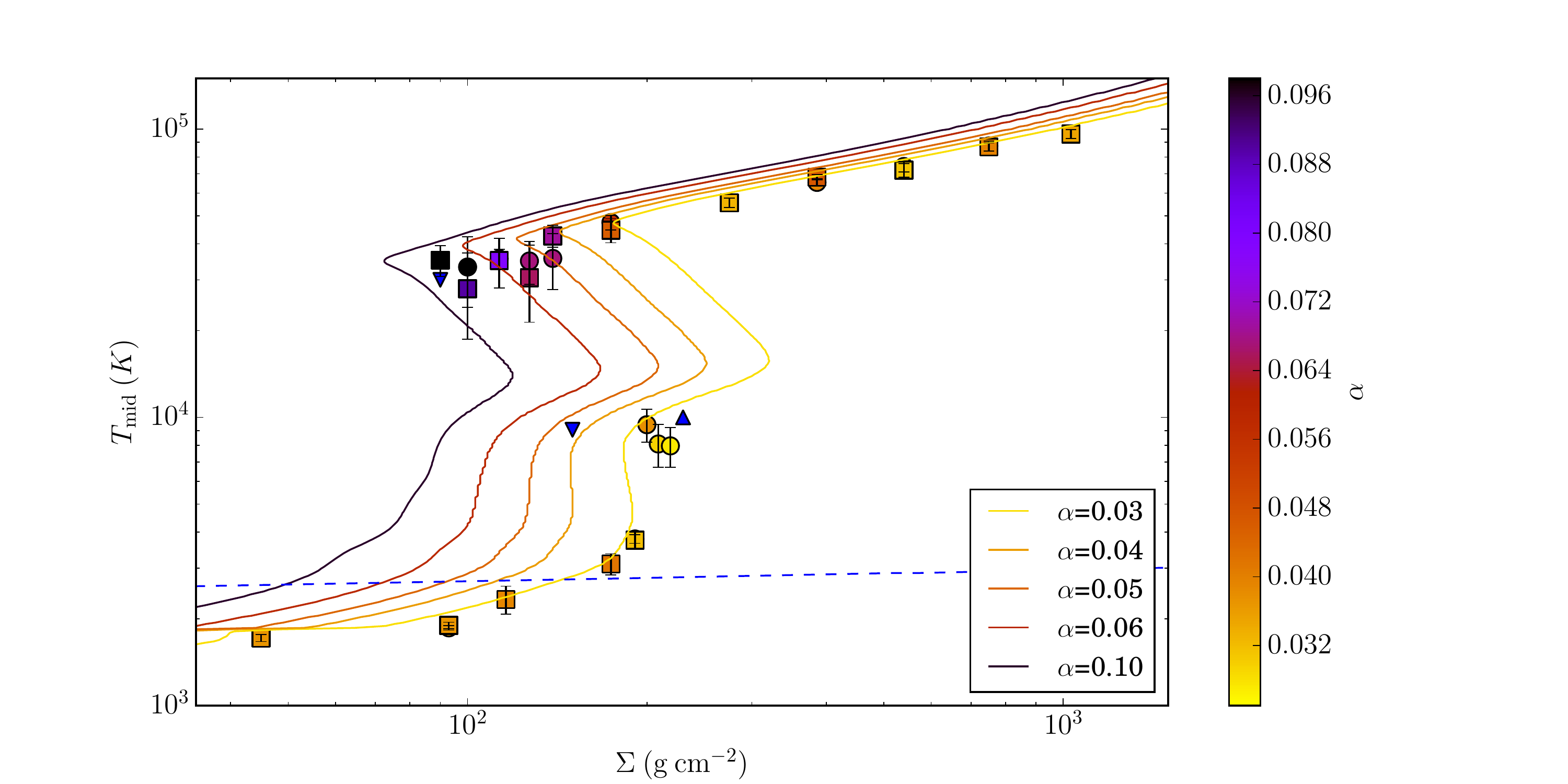}
\includegraphics[width=200mm,height=100mm]{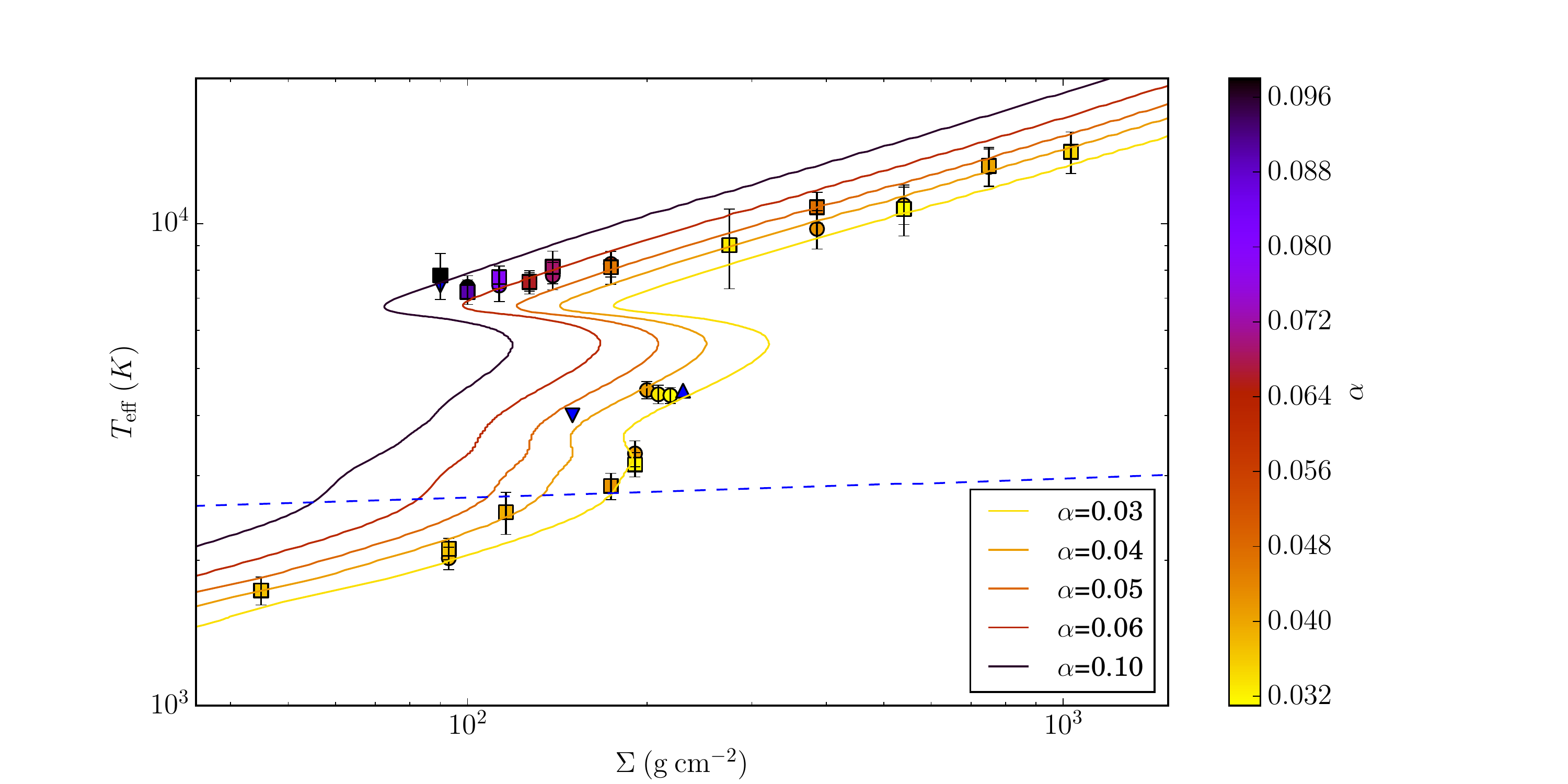}
\end{center}
\caption{Thermal equilibria in the $[T_\mathrm{mid}]$  vs $[\Sigma]$ (top) or  $[T_\mathrm{eff}]$ vs $[\Sigma]$ (bottom)  plane. Squares and circles are respectively for periodic and outflow runs. Triangular dots represent runs with runaway cooling (triangle facing down) or heating (triangle facing up). Error bars represent the standard deviation of the temperature fluctuations. The symbols are color-coded to the value of $\alpha$. The color-coded curves are vertical thermal equilibria using an $\alpha$-prescription. The dashed blue line indicates where the magnetic Reynolds number $R_m=10^4$ based on an isothermal model (see \S\ref{subsec:decay}).}
\label{Scurve}
\end{figure*}

Figure\,\ref{Scurve} shows the thermal equilibria reached by the simulations listed in Table~\,\ref{tab}. The top panel shows the time-averaged midplane temperature $[{T_\mathrm{mid}}]\equiv[\braket{T(z=0)}_{x,y}]$ versus the  surface density $[\Sigma]$ (top panel). The bottom panel shows $[{T_\mathrm{eff}}]$ vs $[\Sigma]$ (bottom panel). The intensity of $\alpha$ is indicated for each run by the color of the dot. The colored curves on Fig.~\ref{Scurve} are thermal equilibria curves obtained  by using the \citet{hameury1998} code, which assumes an $\alpha$-prescription and allows for thermal convection (\S\ref{sec:init}).  Whereas the latter requires choosing a value for $\alpha$, the simulations do not have this degree-of-freedom since the value of $\alpha$ is set by the turbulent angular momentum transport generated by the MRI.

The simulations trace an equilibrium thermal curve in the shape of an S (an S-curve) with a hot, stable branch and a cold, stable branch, providing independent confirmation of the results  of  \cite{Hirose2014}. The S-curve from the simulations is comparable to that predicted by the vertical structure calculations using an $\alpha$-prescription. These also predict a third stable branch at intermediate temperatures ($T_{\rm mid}\approx 10^4\rm\,K$). Indeed, Fig.~\ref{Scurve} shows also a set of three stable runs in an intermediate regime of temperature. A middle-branch in radiative MRI simulations was also reported by \cite{hirose2015} in the context of protoplanetary disks. The middle branch is not as extended as the other branches and may be seen as a prolongation of the cold branch to higher temperatures. In fact, the middle branch does not overlap with the cold branch, so there is at most two equilibrium states for a given $\Sigma$, not three. However, the middle branch runs have a higher opacity than the cold branch and are convectively unstable (\S\ref{sec:conv}).

The hot and cold/middle branch are terminated at, respectively, low and high $\Sigma$ by unstable runs displaying runaway cooling or heating (noted R in Tab.\,\ref{tab} and indicated by arrows in Fig.~\ref{Scurve}). We have followed these runaways to the stable equilibrium on the opposite branch using appropriate box sizes.  Periodic simulations seem to be slightly more stable around points of critical heating/cooling as 453O is unstable whereas 453P is not. Restarting simulations from previous runs by changing the surface density allows us to follow the hot branch to lower surface densities than in \cite{Hirose2014} where 442O is their last stable run.

We have good agreement between our simulations and the $\alpha$-prescription calculation concerning the equilibrium temperatures at given $\alpha$ and the critical points for runaway heating/cooling. The equilibrium values follow the calculated curves corresponding to the value of $\alpha$ found in the simulation. Somewhat unexpectedly, the agreement extends to the middle branch even though its location in the $(\Sigma, T)$ plane depends on the chosen value of the convection mixing length parameter \citep{1984ApJS...55..367C}. Here, we took $\alpha_{ml}=1.5$ as in the solar convection zone (\S\ref{sec:init}).

\subsection{Convection and transport of angular momentum\label{sec:conv}}
\subsubsection{Enhancement of $\alpha$} 
\begin{figure}[h!]
\begin{center}
\includegraphics[width=90mm,height=65mm]{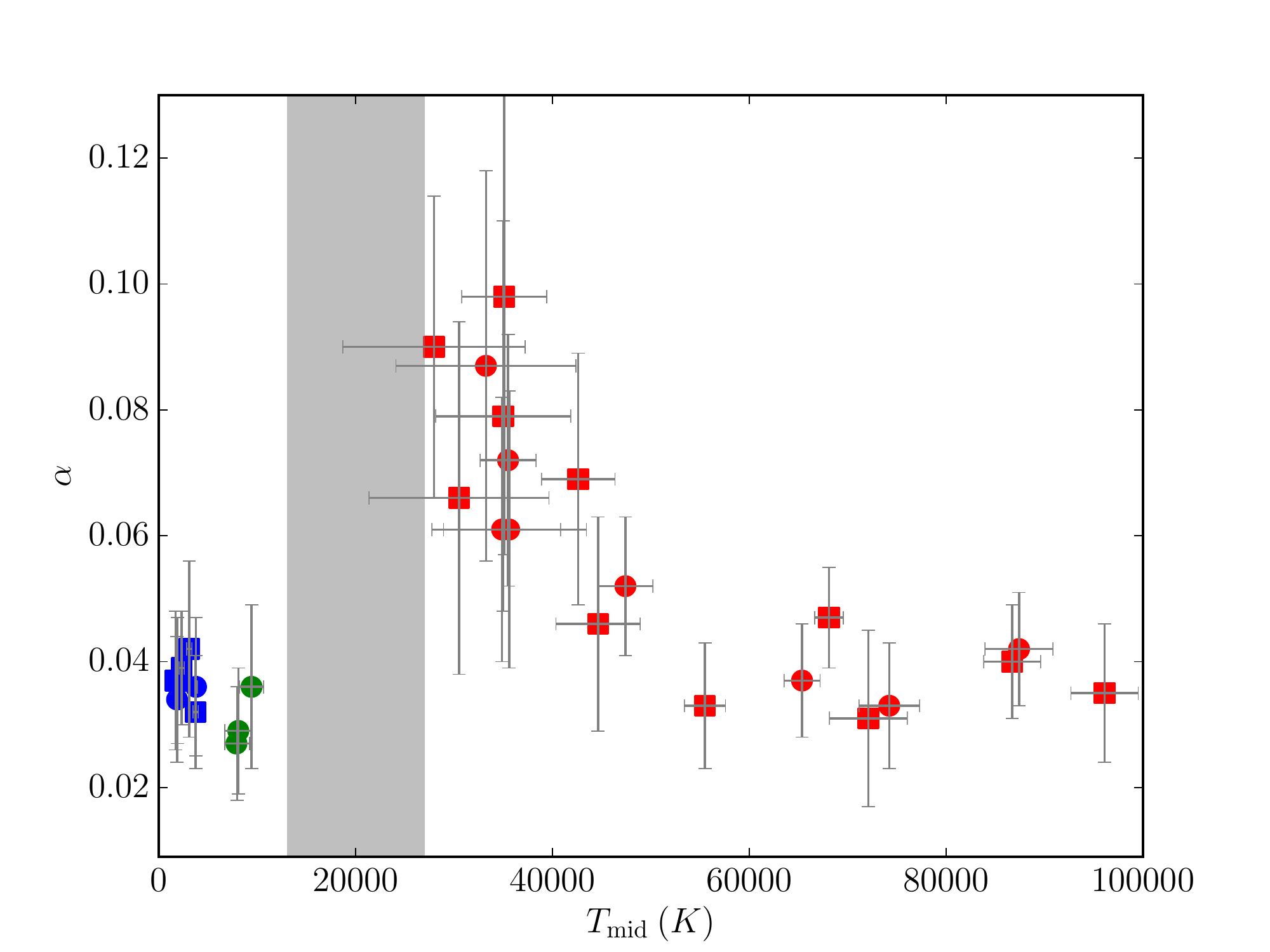}
\includegraphics[width=90mm,height=65mm]{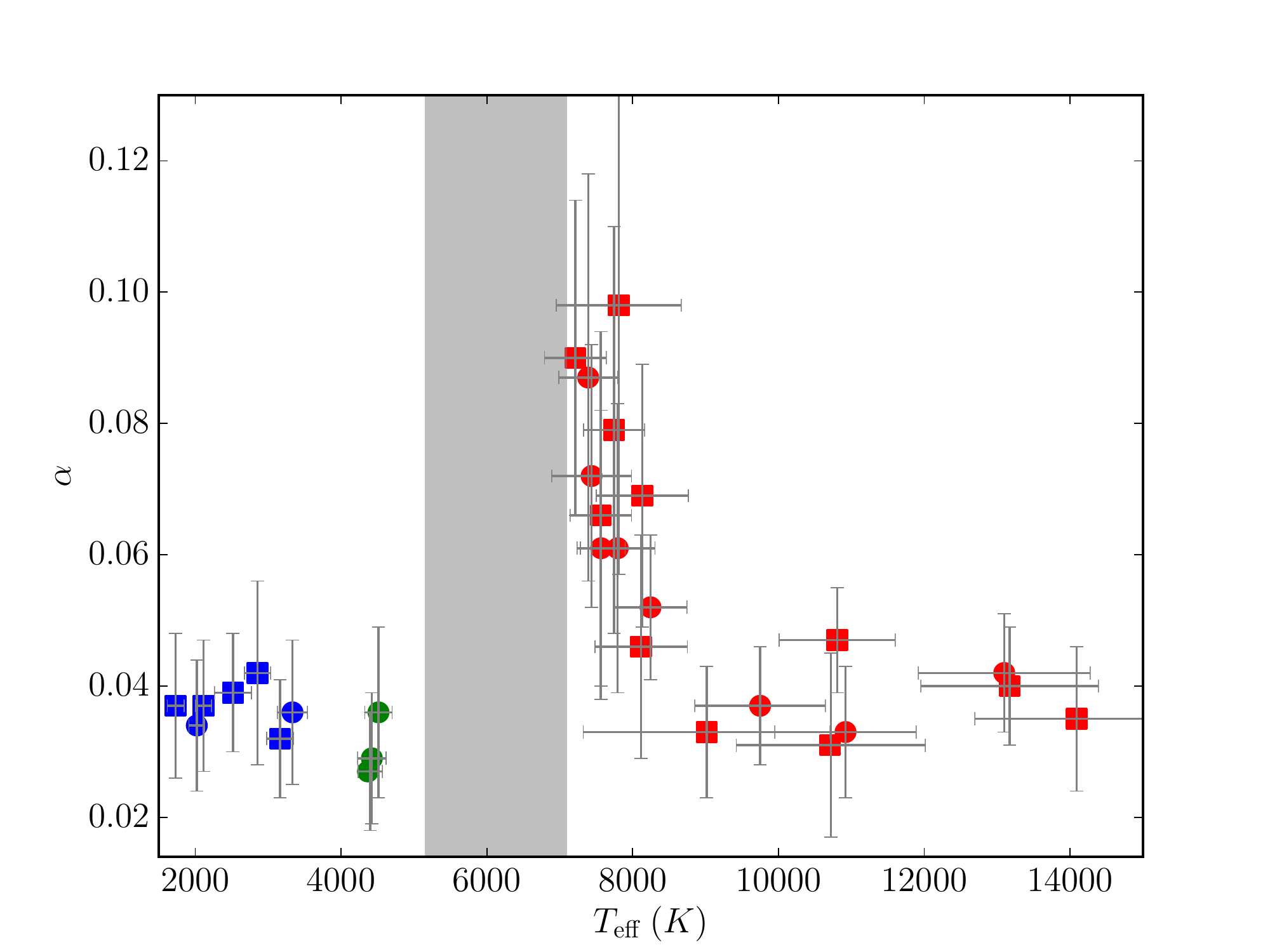}
\end{center}
\caption{$\alpha$ as a function of $[T_\mathrm{mid}]$ (top panel) and  $[T_\mathrm{eff}]$ (bottom panel). Error bars represent the standard deviations of the fluctuations around the mean values. Blue, green and red colors are respectively for the cold, middle and hot branch. Squares and circles indicate periodic and outflow runs. The shaded area corresponds to the thermally-unstable region. }
\label{alpha_Tmid}
\end{figure}

We see an enhancement of $\alpha$ in the low $\Sigma$ part of the hot branch, albeit with slightly lower maximum values of $\alpha\approx 0.098$ than the $\alpha\approx 0.121$ found by \citet{Hirose2014}. On the cold and middle branch, we find values of $\alpha$ ranging from 0.029 to 0.042 typical of a zero net vertical flux stratified MRI simulation \citep{simon2012}. The values of $\alpha$ are comparable on the hot branch for $174\,\mathrm{g\,cm^{-2}}\leq \Sigma \leq 1030\,\mathrm{g\,cm^{-2}}$. For higher $\Sigma$ the value of $\alpha$ increases up to 0.066; we did not further investigate this trend as this part of the S-curve implies very high accretion rates that are unlikely to be relevant to dwarf novae. The value of $\alpha$ also increases as we approach the tip of the hot branch, going to a maximum of 0.098 for the lowest-density run on the hot branch (453P).  This enhancement is clearly seen when plotting $\alpha$ against temperature (Fig.~\ref{alpha_Tmid}, see also Fig.~1 in \citealt{2016MNRAS.462.3710C}). The enhancement is accompanied by stronger fluctuations as we near the unstable tip of the hot branch (see \S\ref{sec:cycle}).

\subsubsection{Convective transport of heat}
 
\citet{Hirose2014} attributed the enhanced $\alpha$ to convection. We also observe that the enhanced $\alpha$ runs are convectively-unstable and further note that the value of $\alpha$ does not depend on the chosen vertical boundary conditions. We measure the stability of a run against convection from  the Brunt-V\"ais\"al\"a frequency $N$, defined by
\begin{equation}
\frac{N^{2}}{\Omega^2}\equiv\frac{d\:\mathrm{ln}\big([\braket{P}]^{1/[\braket{\Gamma}]}[\braket{\rho}]\big)}{d\:\mathrm{ln}z}
\end{equation}
where $\Gamma\equiv(\partial\:\mathrm{ln}\:P/\partial\:\mathrm{ln}\:\rho)_s$ is the adiabatic index and $s$ is the specific entropy. $N^2$ has the sign of the entropy gradient along $z$, thus a negative Brunt-V\"ais\"al\"a frequency denotes a convectively-unstable vertical profile. The Brunt-V\"ais\"al\"a frequency gives a linear stability criterion for the convective stability only in regions where the total pressure is dominated by thermal pressure and does not give direct information about the convection flux. 

\begin{figure}[h!]
\begin{center}
\includegraphics[width=90mm,height=63mm]{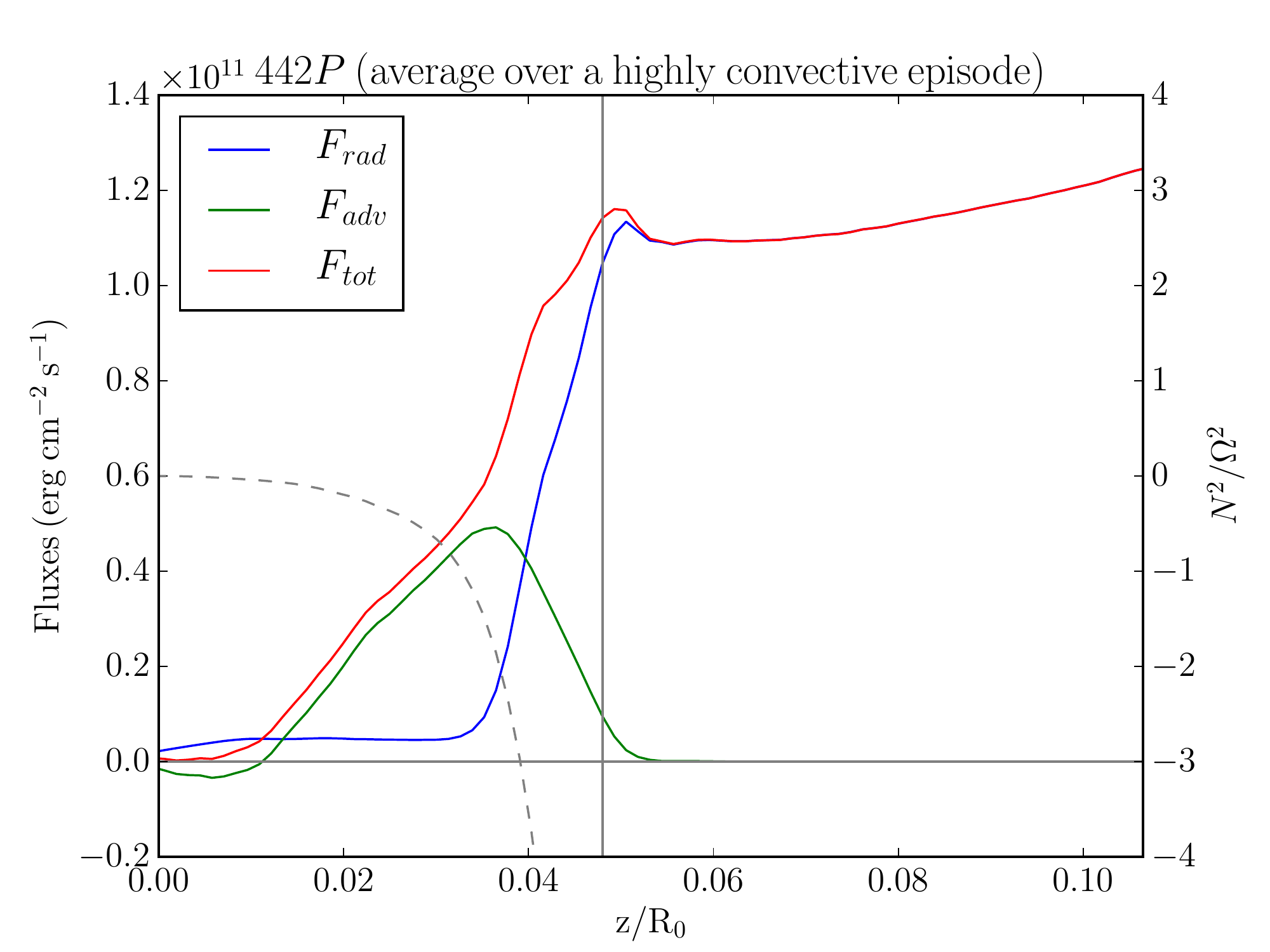}
\includegraphics[width=90mm,height=63mm]{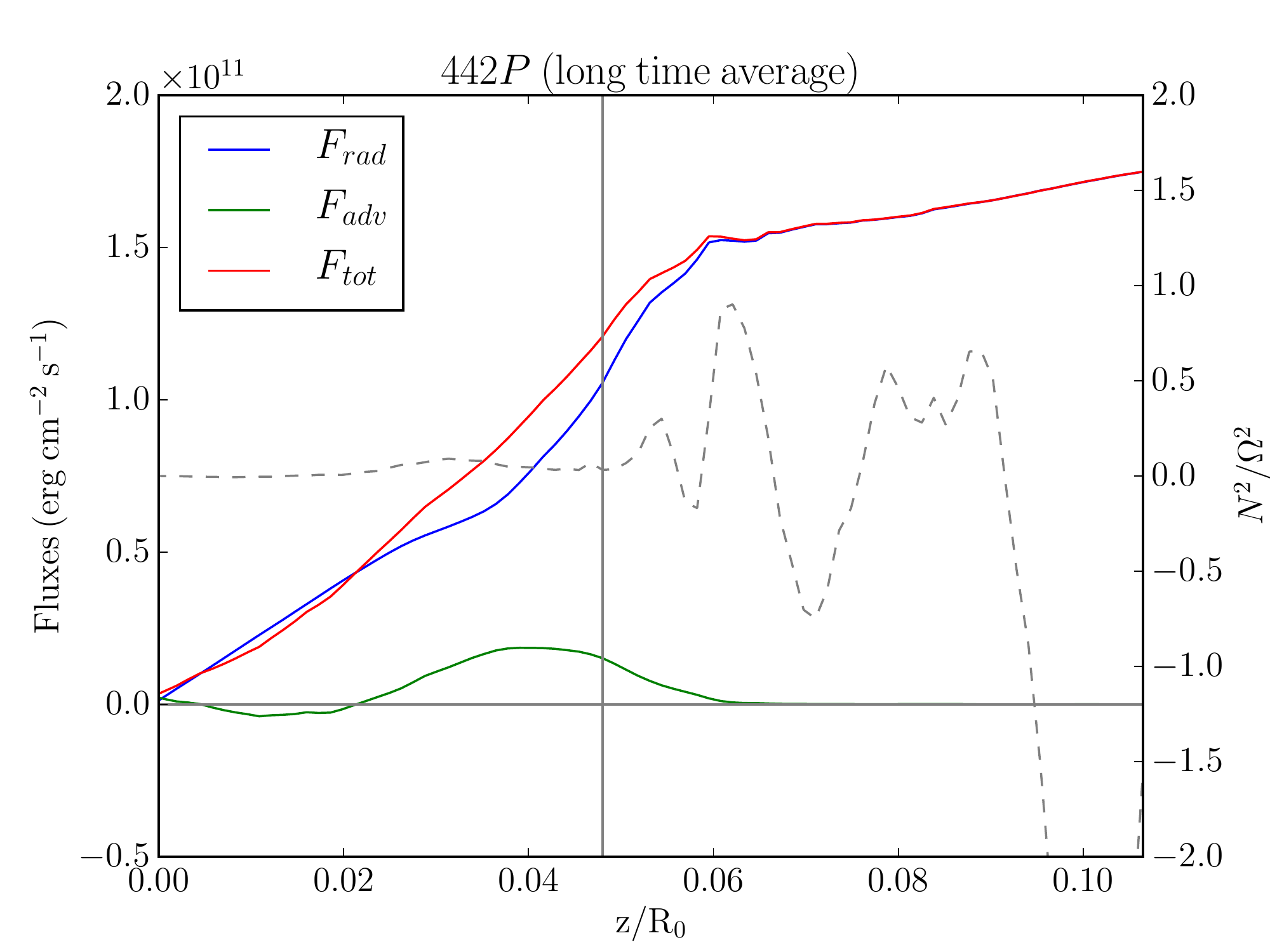}
\includegraphics[width=90mm,height=63mm]{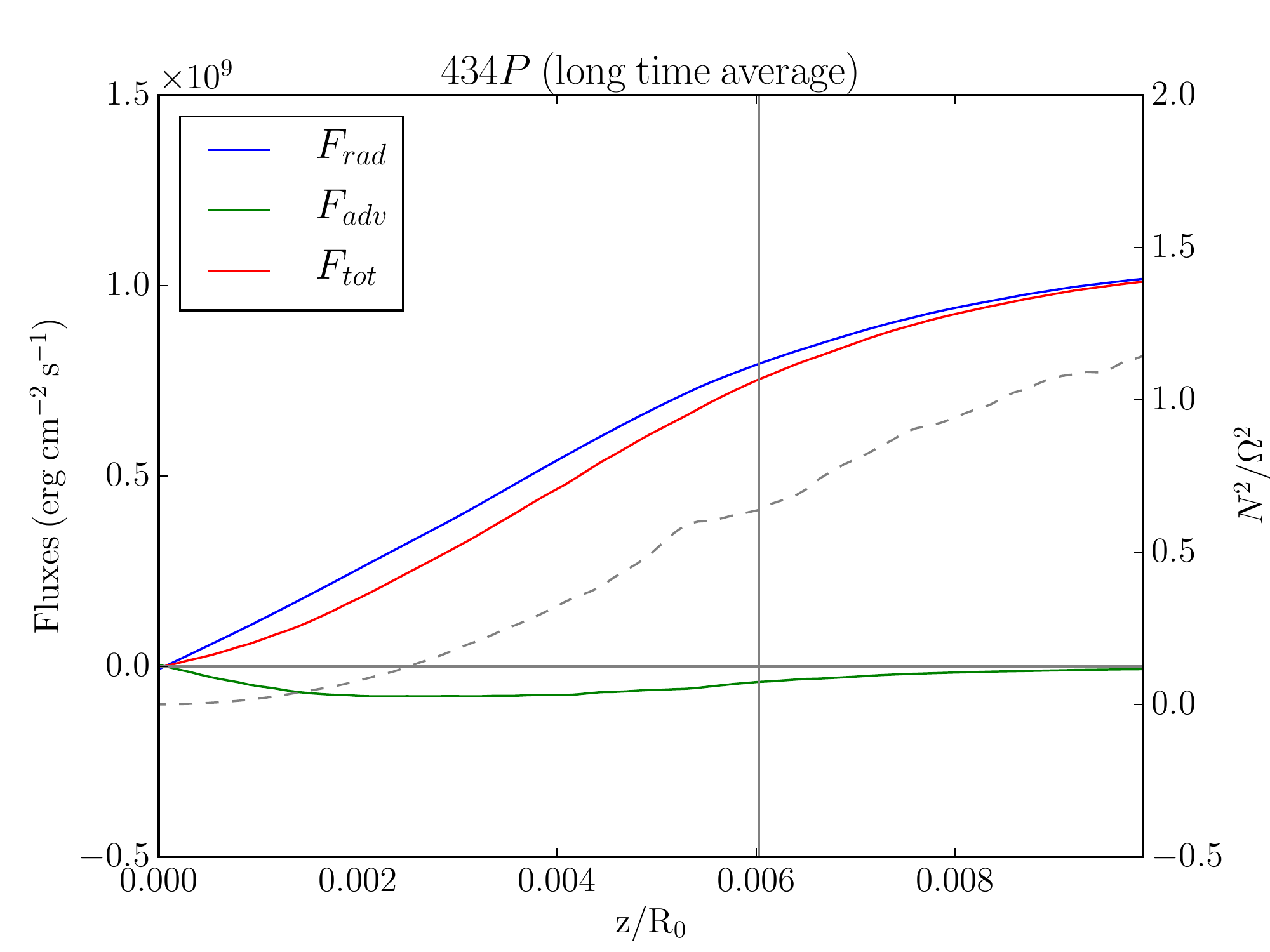}
\end{center}
\caption{Time-averaged vertical profiles of the radiative, advective and total flux for different runs. Top panel: run 442P during a highly convective episode. Middle panel: same run but averaging over convective and non convective episodes. Bottom panel: run 434P where there is no convection. The vertical line indicates the height above which magnetic pressure becomes larger than thermal pressure. The dashed line shows $N^2/\Omega^2$ where $N$ is the Brunt-V\"ais\"al\"a frequency.}
\label{Flux_442F}
\end{figure}

Figure\,\ref{Flux_442F} shows the radiative, advective and total fluxes (Eq.\,\ref{eq:fluxes}) as well as the Brunt-V\"ais\"al\"a frequency as a function of height for several illustrative cases. The top panel shows a highly-convective episode (averaging over 16 local orbits) of the run 442P (hot branch).  The advective flux is the main source of heat transport in the regions where the disk is convectively unstable ($N^2<0$) as opposed to the upper regions where the radiative flux dominates. Averaging the same run on a longer time (80 local orbits) including convective and non-convective episodes shows the contribution of the advective flux decreases (middle panel). We also see  evidence for downward transport of heat between the midplane and $H/R_0 \sim 2.3\times10^{-2}$. This downward transport is also present in the top panel but  is less extended in height. Convection is expected to drive heat upward, where the entropy is lower, hence this downward transport is not due to convective motions. In fact,  run 434F on the cold branch is convectively stable yet also shows this downward advective transport (bottom panel). We attribute this transport to vertical mixing by the MRI-driven turbulence. Because motions are approximately adiabatic in this region, this mixing tends to flatten the entropy profile, resulting in a downward transport of heat. Thus, the advective flux includes contributions from both convective motions (when present) and turbulent mixing. 

Besides the region of enhanced $\alpha$ near the unstable tip of the hot branch, we find convection also plays a major role in the middle branch runs, notably run 441O, with the fluxes behaving as in the top panel of Fig.~\ref{Flux_442F}.

\subsubsection{Convective/Radiative cycles\label{sec:cycle}}

In the following, we make use of the convective fraction $f_\mathrm{conv}$ of the flux, defined as
\small
\begin{equation}
f_\mathrm{conv} \equiv \frac{1}{2}(\mathrm{max}(0,\mathrm{max}(\big[\frac{F_\mathrm{adv}(z>0)}{F_\mathrm{rad\:z}^+}\big]))+  \mathrm{max}(0,\mathrm{max}(\big[\frac{F_\mathrm{adv}(z<0)}{F_\mathrm{rad\:z}^-}\big])))
\end{equation}
\normalsize
We also use an instantaneous convective fraction $\tilde{f}_\mathrm{conv}$ obtained by smoothing over a time window with a width of 1.6 local orbits. We note that $\tilde{f}_\mathrm{conv}$ can have values $>1$  due to the  fluctuations remaining in the vertical profile.

\begin{figure}[h!]
\begin{center}
\includegraphics[width=90mm,height=65mm]{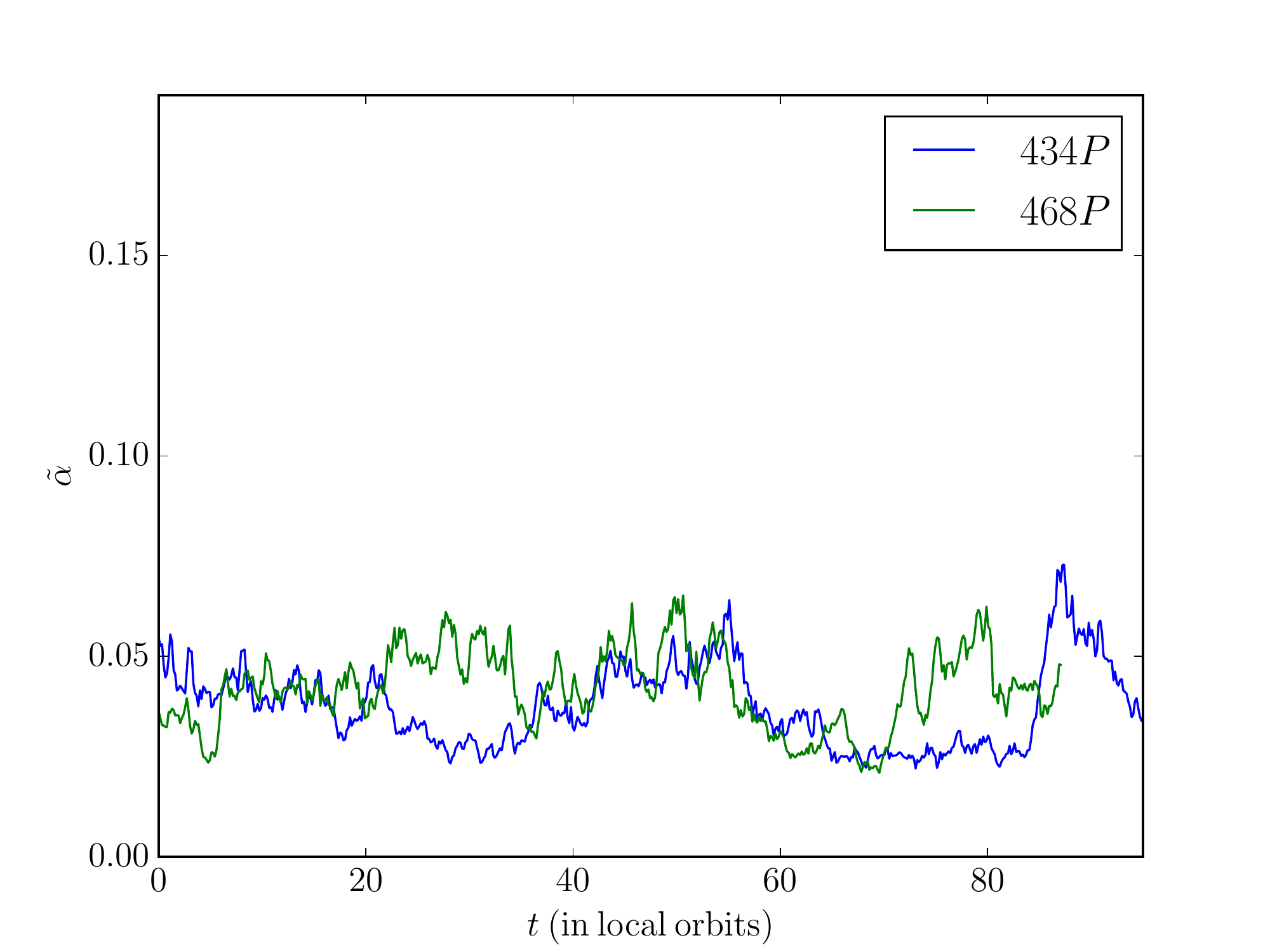}
\includegraphics[width=90mm,height=65mm]{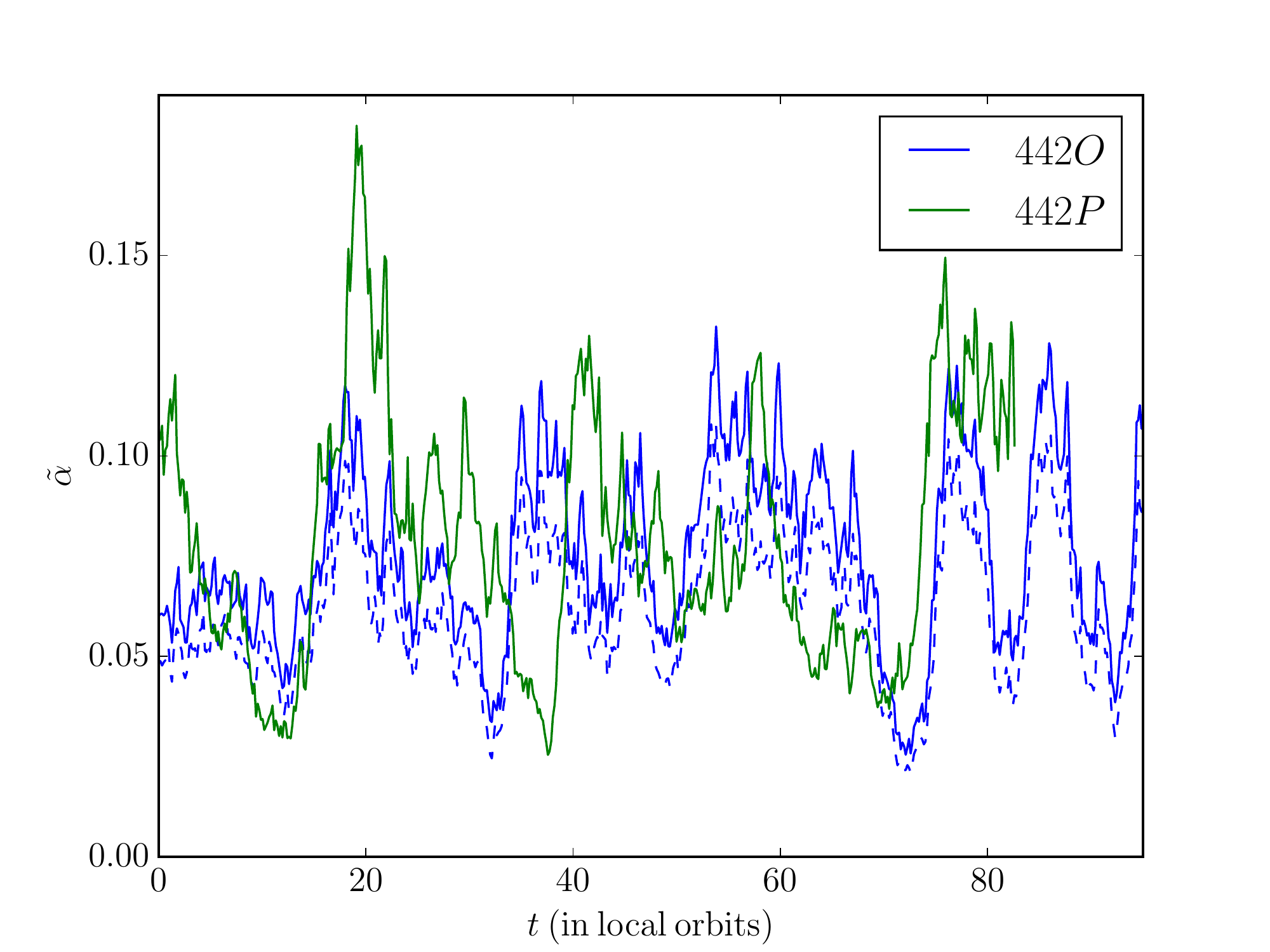}
\end{center}
\caption{Evolution of $\tilde{\alpha}$ as a function of time for runs located on different parts of the S-curve: (top panel) 434P is a cold branch run, 468P a non-convective hot branch run, and (bottom panel) 442 a convective hot branch run. For the latter we show the simulation ran with outflow  (442O) and periodic boundary conditions (442P). We also plot, for the run 442O, the contribution of the Maxwell  stress to $\tilde{\alpha}$ as a dashed line.}
\label{alpha_434_468_442}
\end{figure}

Figure \ref{alpha_434_468_442} shows the evolution of $\tilde{\alpha}$ with time for runs 434P, 468P, 442P and 442O.  The first two are convectively-stable and show only weak fluctuations in $\tilde{\alpha}$. However, run 442 is located near the tip of the hot branch and is convectively-unstable. This runs shows $\tilde{\alpha}$ has large fluctuations, going through cycles with bursts of strong angular momentum transport. We see from Figure \ref{alpha_434_468_442} that the main contribution to $\tilde{\alpha}$ is the Maxwell part. The maximum and the recurrence of these bursts in $\tilde{\alpha}$ increase towards the unstable tip of the S-curve. For comparison, there is one cycle every $\sim$20 orbits in run 437O ($\Sigma=174\rm\,g\,cm^{-2}$) and one every $\sim$10 orbits in run 442O ($\Sigma=113\rm\,g\,cm^{-2}$).

\begin{figure}
\begin{center}
\includegraphics[width=90mm,height=60mm]{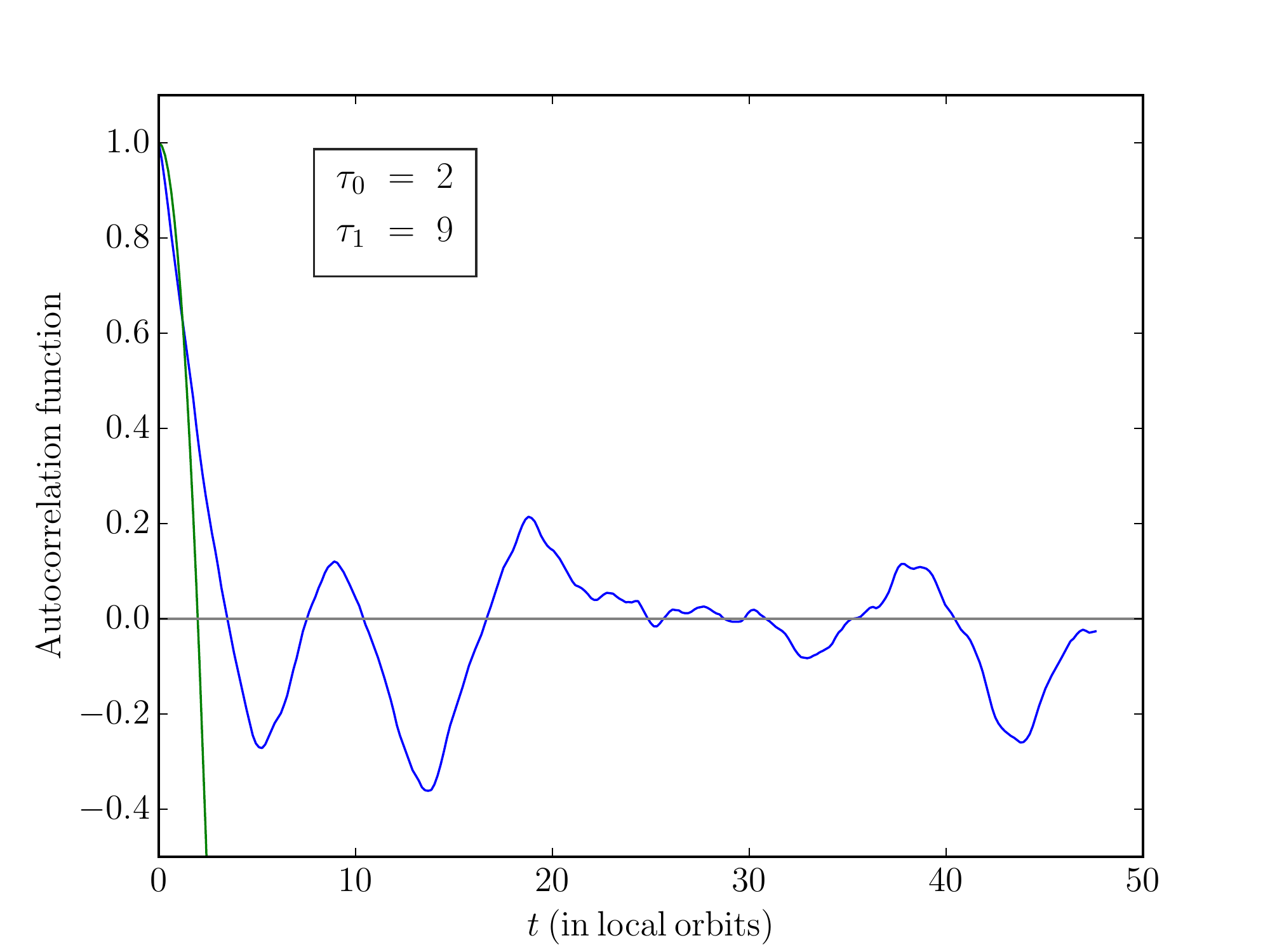}
\end{center}
\caption{Autocorrelation function  $f_{\rm auto}$ as a function of time for the run 442O. $\tau_0$ is the parabolic decay time of the autocorrelation function (with the parabolic fit shown as a dashed line. $\tau_1$ is the location of the first peak of $f_{\rm auto}$. }
\label{autocorr}
\end{figure}

To be more quantitative about the frequency of the cycles of $\tilde{\alpha}$, we compute its autocorrelation function
\begin{equation}
f_\mathrm{auto}(\tau) = \frac{\int(\tilde{\alpha}(t)-\alpha)(\tilde{\alpha}(t+\tau)-\alpha) \mathrm{dt}}{\int(\tilde{\alpha}(t)-\alpha)^2 \mathrm{dt}}
\end{equation}
Figure \ref{autocorr} shows this function for run 442O. We extract two characteristic times. The first is the parabolic decay time $\tau_0$ of $f_{\rm auto}$ ($\tau_0\approx 2$ local orbits in Fig.~\ref{autocorr}. It corresponds to the time on which small fluctuations of $\tilde{\alpha}$ remain correlated. The second, $\tau_1$, corresponds to the first peak of $f_{\rm auto}$ ($\tau_1\approx 9$ local orbits in Fig.~\ref{autocorr}). It corresponds roughly to the period of the convective cycles. 

\begin{figure}
\begin{center}
\includegraphics[width=90mm,height=60mm]{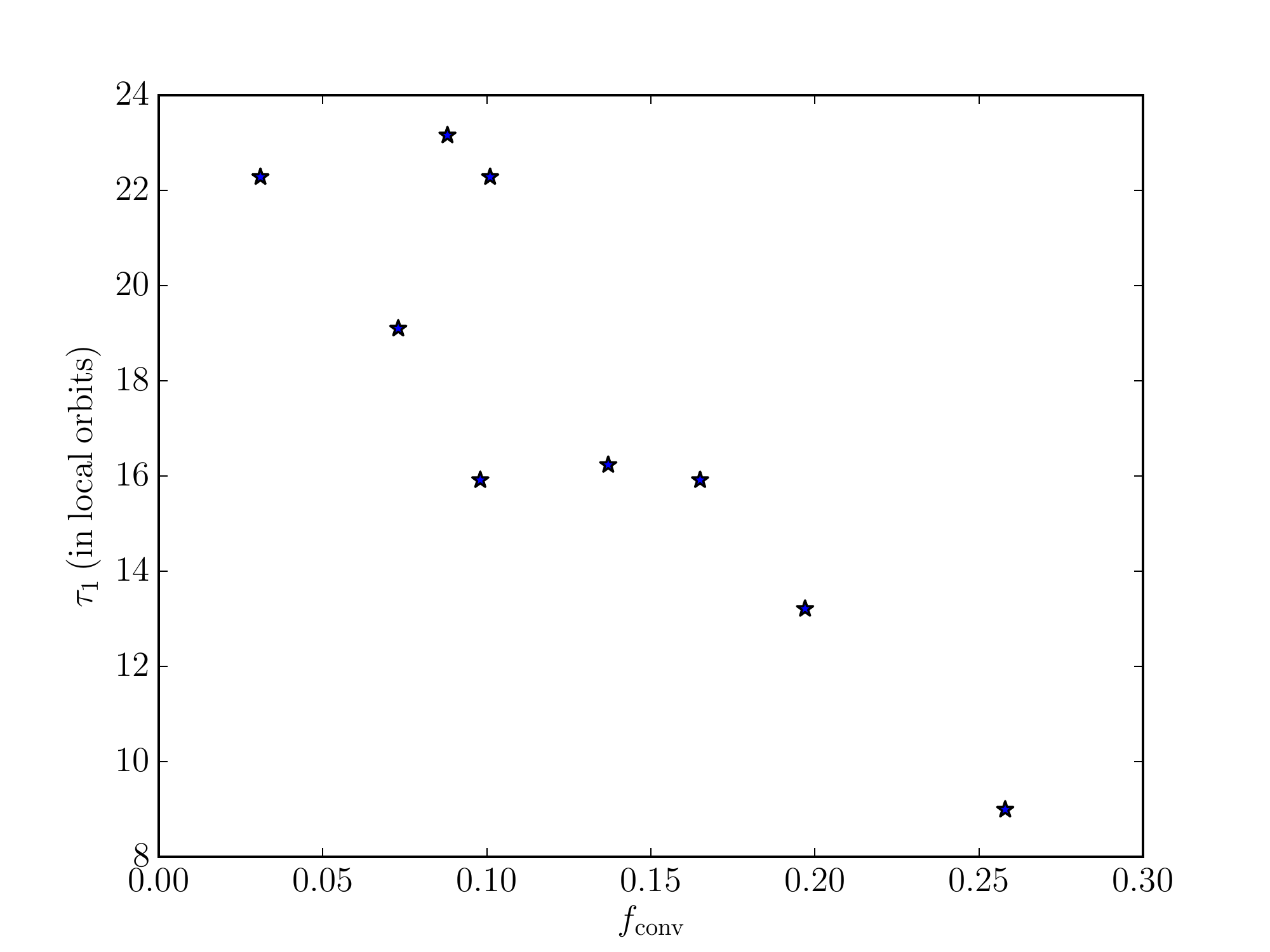}
\end{center}
\caption{Period $\tau_1$ of the cycles in $\alpha$ as a function of the convective fraction for convectively-unstable simulations on the hot branch.}
\label{tau1}
\end{figure}

We do not see any clear evolution of $\tau_0$  as the convective fraction increases on the hot branch. However, $\tau_1$ clearly decreases with $f_{\rm conv}$ (Fig.~\ref{tau1}). On the middle branch, the cycles are much longer for the same $f_{\rm conv}$. For instance, run 401O has $f_{\rm conv}=0.14$ and $\tau_1\approx 111$ local orbits, requiring a longer averaging timescale $t_{\rm avg}$ than for the other runs. 

\begin{figure}
\begin{center}
\includegraphics[width=85mm,height=60mm]{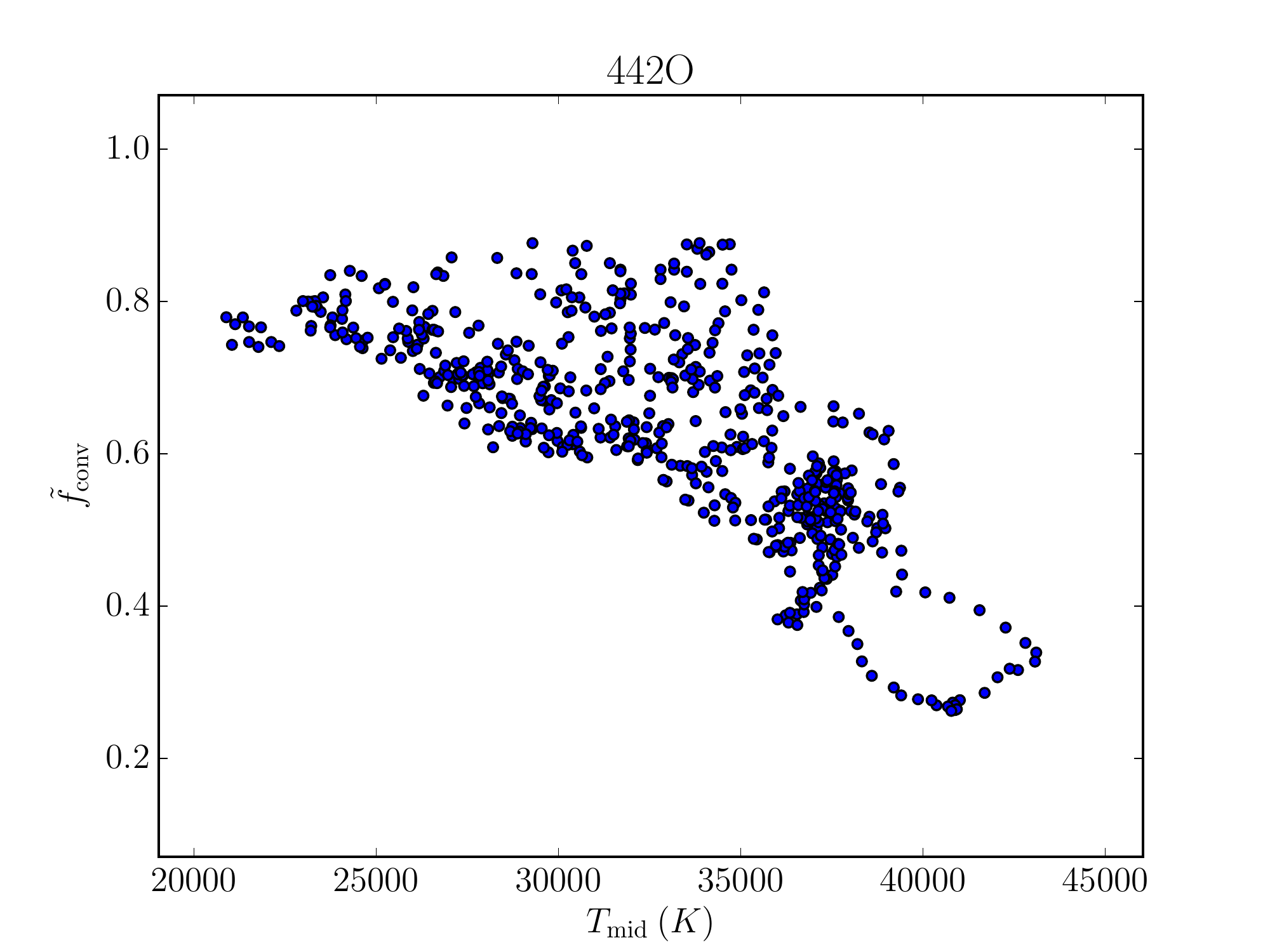}
\includegraphics[width=85mm,height=60mm]{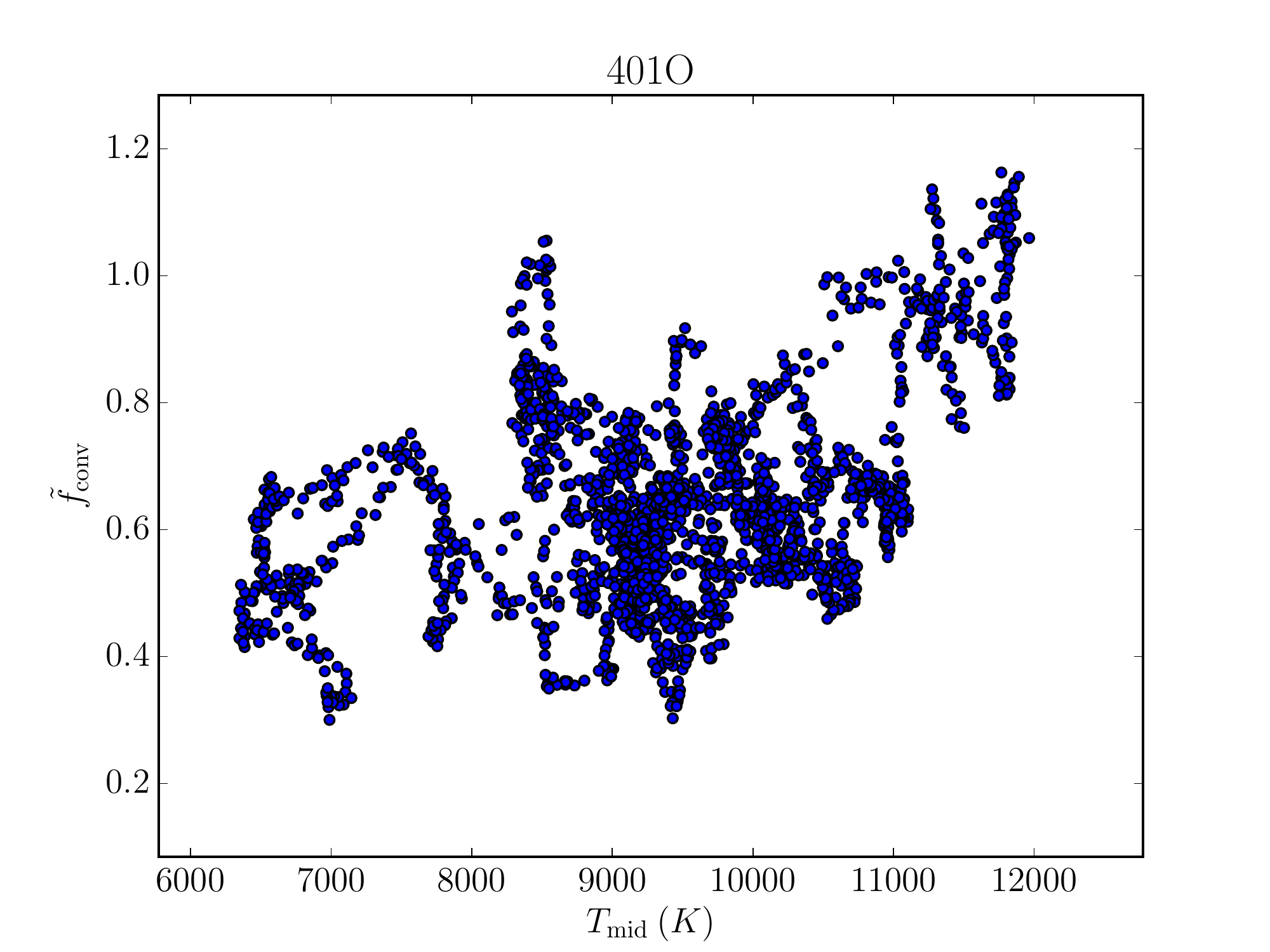}
\end{center}
\caption{$\tilde{f}_\mathrm{conv}$ vs $T_\mathrm{mid}$ for two representative convectively-unstable simulations:  hot branch run 442O (top) and middle branch run 401O (bottom).}
\label{cycles}
\end{figure}

The stronger variability near the unstable tip of the hot branch is also manifest as increased temperature fluctuations (error bars in Fig.~\ref{Scurve} and Fig.~\ref{alpha_Tmid}). These fluctuations in temperature are anti-correlated with $\tilde{f}_\mathrm{conv}$ on the hot branch (Fig.~\ref{cycles}, top panel). However, the fluctuations are in phase at the highest temperatures on the middle branch (Fig.~\ref{cycles}, bottom panel). The behaviour on the hot and middle branch is very different although the fractional amplitudes ([max-min]/[max+min]) of the fluctuations in $\tilde{f}_\mathrm{conv}$ and $T_{\rm mid}$ are the same for both.  This difference can be explained by noting that the temperature gradient, assuming only radiative transport, is $\propto \kappa/T^3$ and that convection requires this gradient to be greater than the adiabatic gradient of the gas. On the hot branch, the  opacity decreases when the temperature rises, flattening the temperature profile and quenching convection. Upward fluctuations in temperature lead the disk from a convective regime to a radiative regime and vice-versa. The opposite correlation is expected on the middle branch since the opacity rises steeply with $T$, triggering a convective regime when fluctuations raise the temperature. 
\begin{figure}
\begin{center}
\includegraphics[width=85mm,height=60mm]{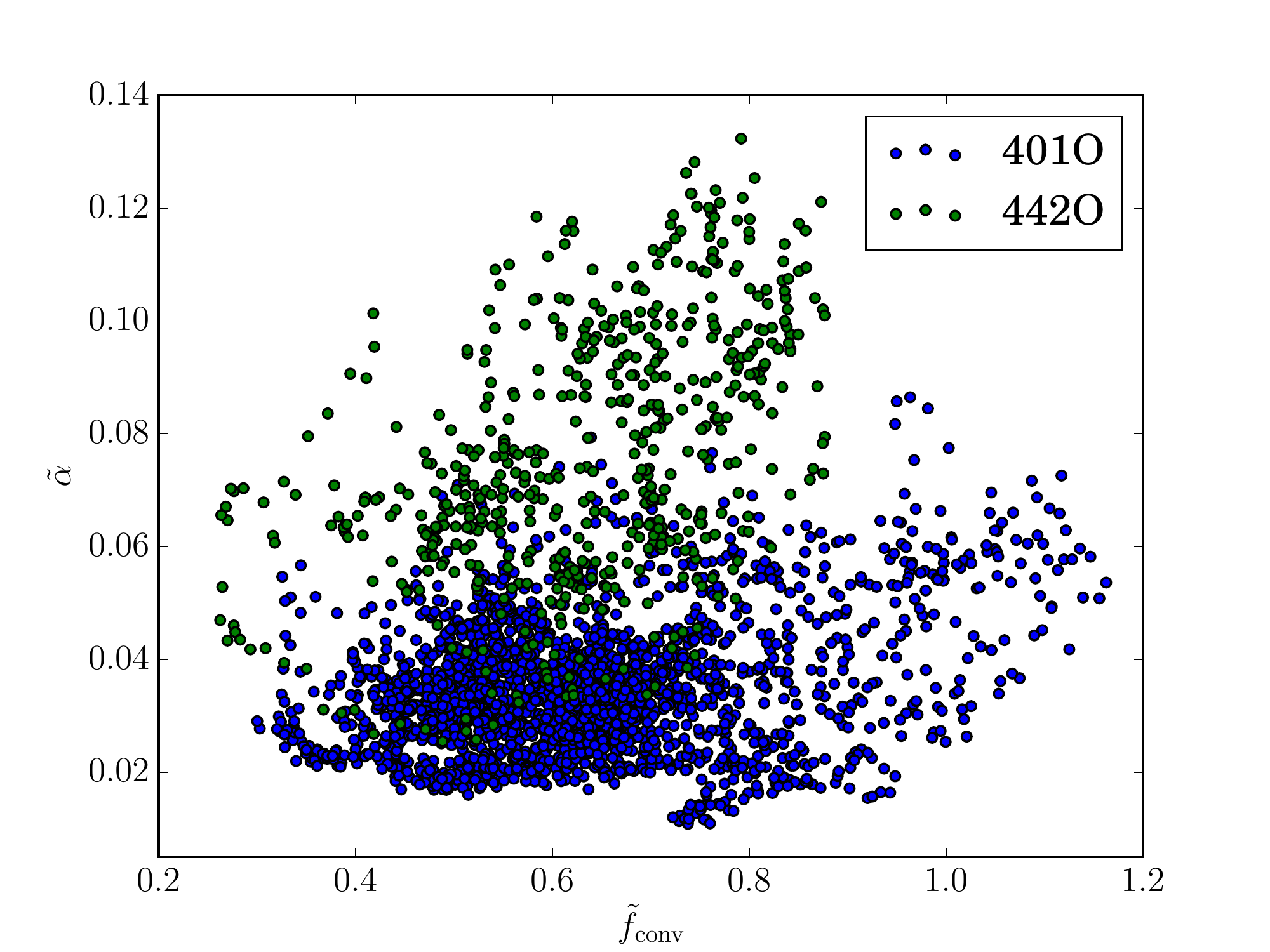}
\end{center}
\caption{$\tilde{\alpha}$ vs $\tilde{f}_\mathrm{conv}$ for the same runs as Fig.~\ref{cycles}.}
\label{cycles2}
\end{figure}

\begin{figure}[h!]
\begin{center}
\includegraphics[width=85mm,height=60mm]{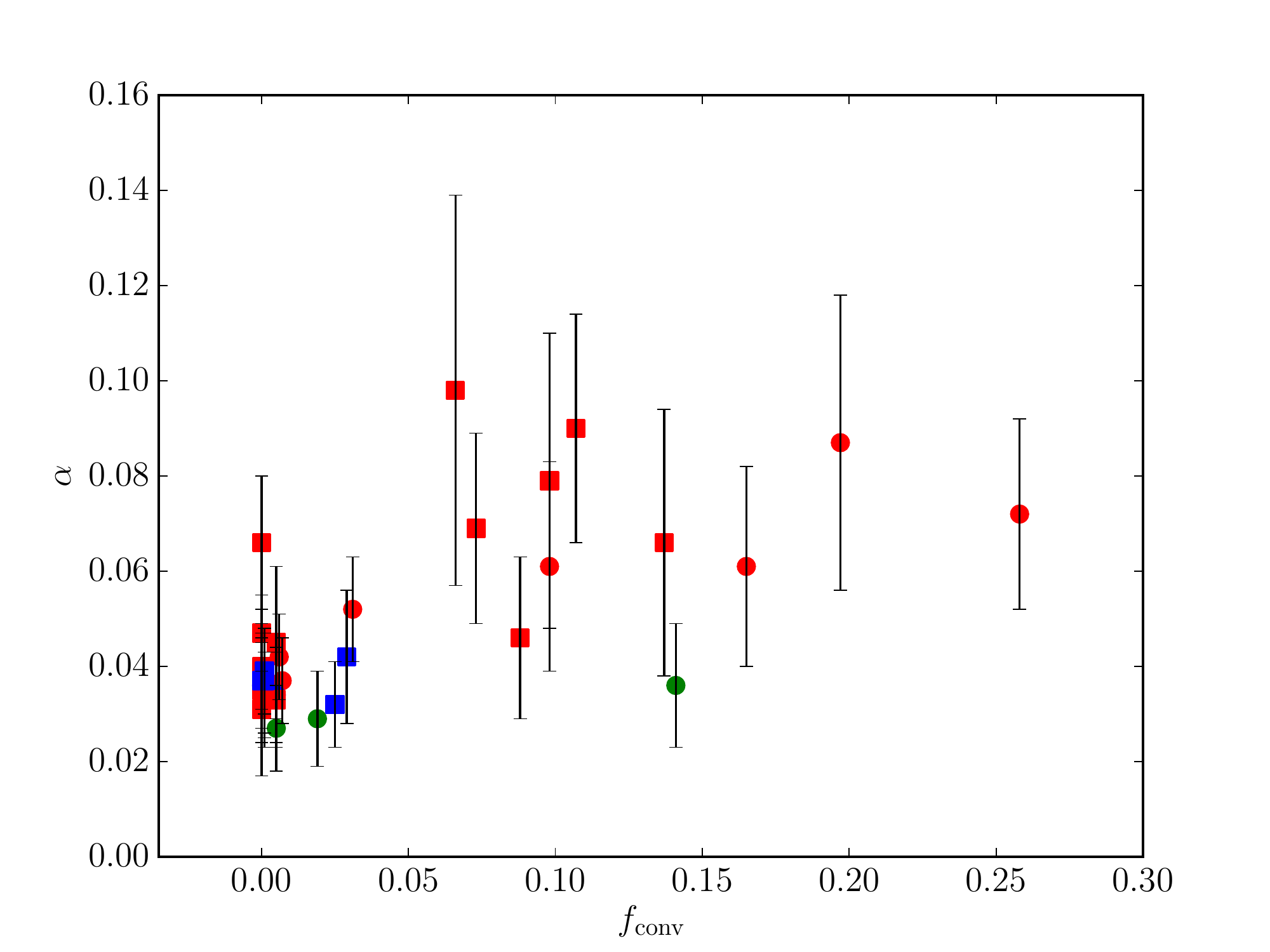}
\end{center}
\caption{$\alpha$ vs convective fraction $f_\mathrm{conv}$ for all convectively-unstable runs (same color and symbol coding as Fig.~\ref{alpha_Tmid}).}
\label{alpha_fconv}
\end{figure}

This difference in behaviour between the convective parts of the hot branch and the middle branch also extends to $\alpha$. Run 442O on the hot branch shows a trend for the instantaneous $\tilde{\alpha}$ to increase with the instantaneous convective fraction  $\tilde{f}_{\rm conv}$ (Fig.~\ref{cycles2}, green markers), but this trend is absent in run 401O on the middle branch (Fig.~\ref{cycles2}, blue markers). Furthermore, the maximum value of $\tilde{\alpha}$ is lower than on the hot branch run even though $\tilde{f}_{\rm conv}$ reaches higher values. The runs also show a difference in duty cycle. The hot branch run spends roughly equal time in the low and high $\tilde{f}_{\rm conv}$ parts of the diagram. In contrast, the middle branch run spends most of its time in the  low $\tilde{f}_{\rm conv}$ part of the diagram, with only temporary excursions to the higher values. This will naturally lead to a smaller time-averaged $\alpha$ in the middle branch. Fig.~\ref{alpha_fconv} shows no trend in the convective runs between the time-averaged values of $\alpha$ and $f_{\rm conv}$. We conclude that there is no simple relationship between the enhancement of $\alpha$  and the relative strength of the convection measured by $f_{\rm conv}$. Finding the physical origin for this enhancement is likely to require a detailed study of the impact of vertical convection on the MRI dynamo, which is beyond the scope of this paper.

\subsection{Impact of the vertical boundary conditions}
\begin{figure}[h!]
\begin{center}
\includegraphics[width=75mm,height=50mm]{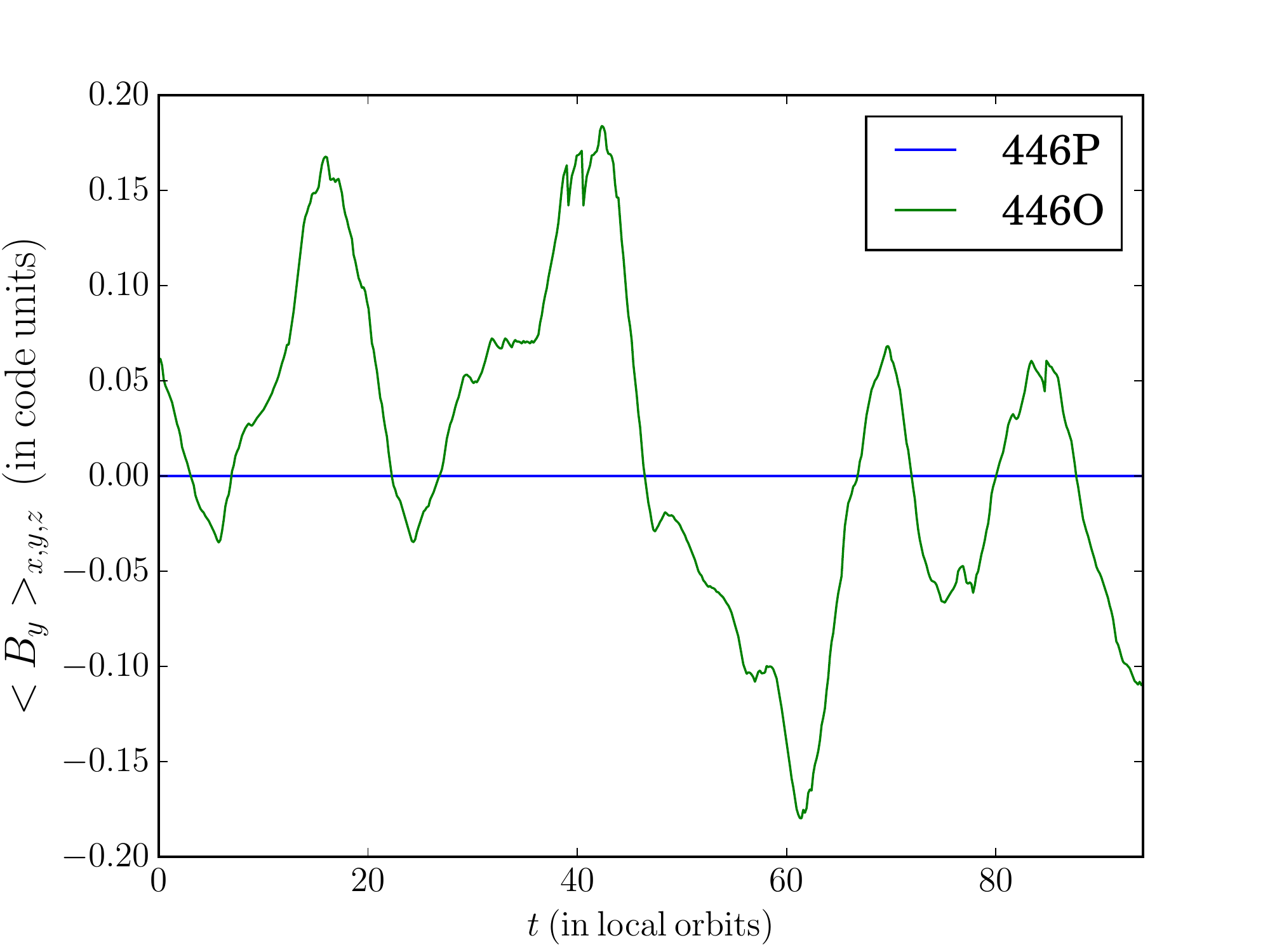}
\end{center}
\caption{Average azimuthal magnetic field $\braket{B_y}_{x,y,z}$ as a function of time  for run 446 (outflow O or periodic P boundary conditions).}
\label{By_t}
\end{figure}

We tested outflow and periodic vertical boundary conditions  (\S\ref{sec:init}) because of their potential impact on the interplay between the MRI dynamo and convection.   In periodic runs, the box-averaged $\braket{B_y}_{x,y,z}=0$ by construction whereas it fluctuates over time with outflow boundary conditions (Fig.~\ref{By_t}). The flips in sign are related to the dynamo of stratified MRI turbulence and is still poorly understood \citep{Brandenburg1995}. We found no quantitative difference between periodic and outflow runs so these flips have no impact on $\alpha$ or on the thermal equilibrium values.  

\begin{figure}[h!]
\begin{center}
\includegraphics[width=100mm,height=45mm]{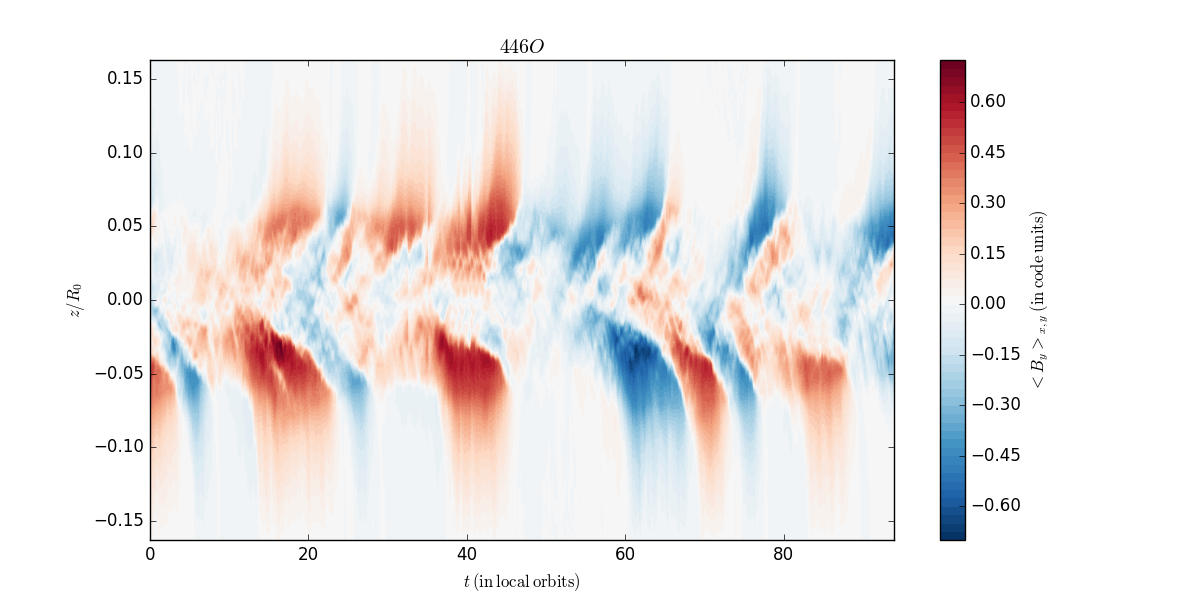}
\includegraphics[width=100mm,height=45mm]{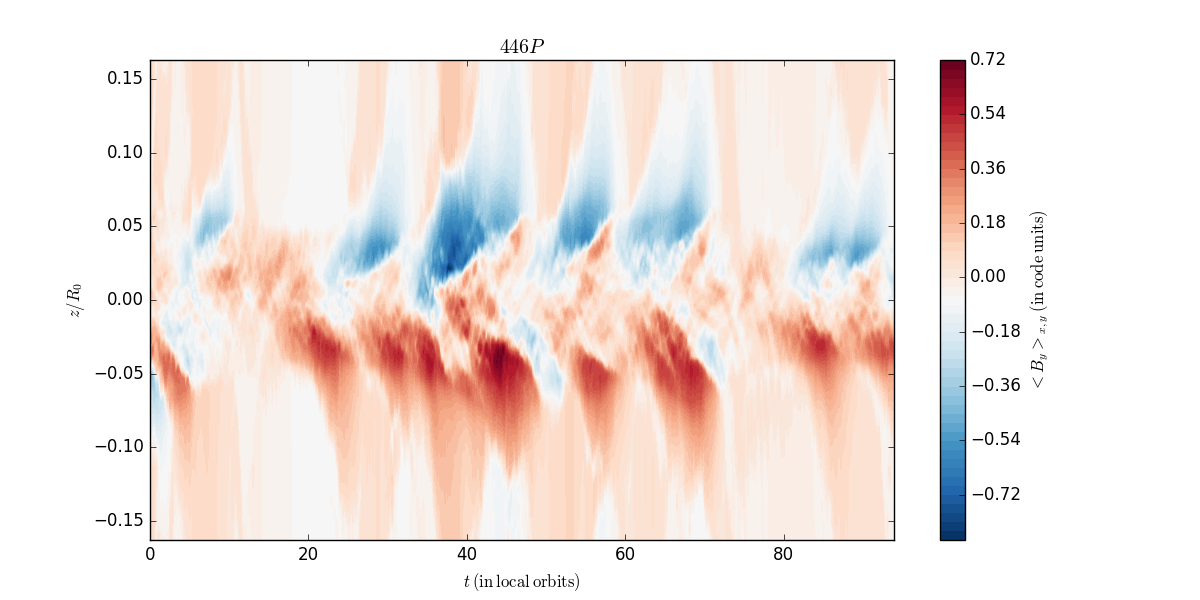}
\end{center}
\caption{Vertical profile of azimuthal magnetic field $\braket{B_y}_{x,y}$ (in code units) as a function of time with outflow (top) and periodic (bottom) boundary conditions for the run 446 (hot branch).}
\label{By_locking}
\end{figure}

The similar behaviour between outflow and periodic runs extends to the "butterfly" diagram that emerges from maps of the time evolution of $\braket{B_y}_{x,y}$, the vertical profile of the azimuthal field. Non-convective MRI simulations are known to form "butterfly wings" as the magnetic field flips sign at each quasi-periodic pulsation and propagate outward (e.g. see Fig.~\ref{435F}). \citet{coleman2017} reported that the quasi-periodic pulsation does not always lead to a sign flip of $\braket{B_y}_{x,y}$ when convection is active. We confirm that this is also the case in our convective outflow simulations (Fig.~\ref{By_locking}, top panel) and in our periodic boundary simulations (bottom panel), even though the magnetic field must adopt a symmetric configuration in the periodic simulations to ensure that $\braket{B_y}_{x,y,z}=0$. 
We also see this behavior in the middle branch during convective episodes.

\section{Resistive MHD runs}
Ideal MHD may not apply to the cold branch due to the very low ionization fractions. The development of the MRI can be suppressed as the resistivity due to electron-neutral collisions increase. This is quantified by the magnetic Reynolds number
\begin{equation}
R_\mathrm{m}\equiv\frac{c_s h}{\eta}
\end{equation}
with $\eta$ the resistivity, $c_{s}$ the sound speed and $h=c_{s}/\Omega$ the local pressure scale-height. For $R_{\rm m}<10^4$, it is expected that diffusion of the magnetic field becomes too important for the disk to sustain MHD turbulence \citep{Hawley1996,Fleming2000}. \citep{Gammie1998} pointed out that $R_\mathrm{m}$ is of the order of $10^3$ in a quiescent DNe disk.

We investigate whether our cold branch runs can maintain turbulence when resistivity is included. The resistivity is computed as in \cite{blaes1994}:
\begin{equation}
\eta=230\left(\frac{n_n}{n_e}\right)T^{1/2}\:\mathrm{cm^2.s^{-1}}
\end{equation}
with the assumption that we are dominated by electron-neutral collision. The values of $\eta$ are pre-computed with the equation of state (Saha equilibrium) and saved in a table.  The time step $\Delta t$ associated with the diffusive term is of the order of $\Delta x^2/\eta$ where $\Delta x$ is the cell minimum length. To avoid dramatically small time steps, we impose a minimum value of $R_\mathrm{m}=50$ in our runs. This floor is largely under the limit for MRI turbulence to be suppressed and does not affect our results. We find $\Delta t\sim 10^{-3}$ of the order of the time step of the ideal MHD run. The resistive runs are restarted from the periodic runs in ideal MHD and labelled PR to differentiate them.

\subsection{Turbulence decay due to resistivity\label{sec:decay}}
\label{subsec:decay}
\begin{figure*}[ht!]
\begin{center}
\includegraphics[width=0.7\linewidth]{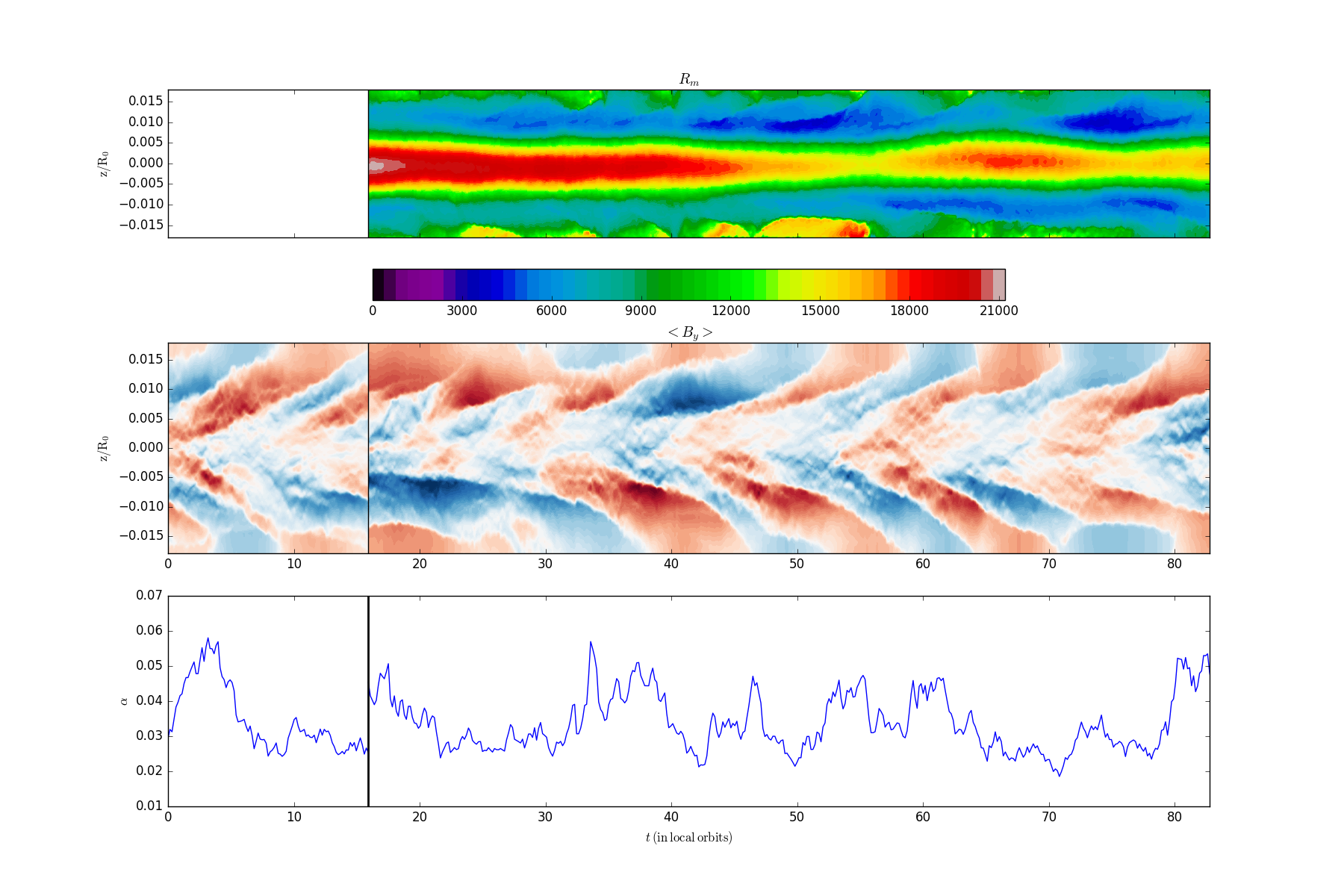}
\end{center}
\caption{Time evolution of $\braket{R_m}$, $\braket{B_y}$ and $\alpha$ for run 435PR ($\Sigma=191\,\mathrm{g\,cm^{-2}}$). The vertical black line marks the time at which the simulation is restarted from 435P with resistivity turned on.}
\label{435F}
\end{figure*}

\begin{figure*}[h!]
\begin{center}
\includegraphics[width=0.7\linewidth]{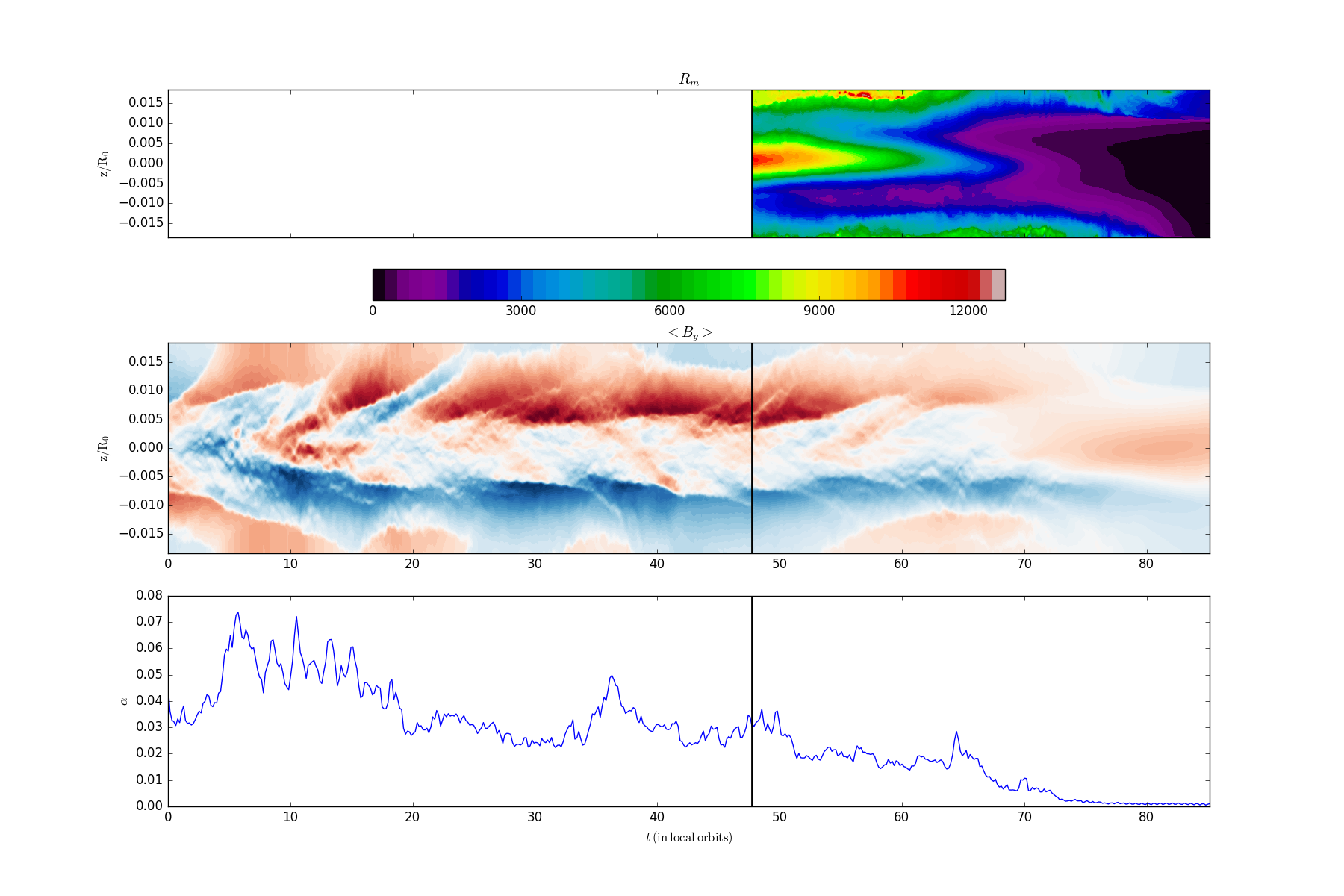}
\end{center}
\caption{Same as Fig.~\ref{435F} for run 462PR ($\Sigma=174\,\mathrm{g\,cm^{-2}}$).}
\label{462F}
\end{figure*}

We find that the resistivity has a critical impact by suppressing turbulence on the cold branch below some critical density located between $\Sigma=174\,\mathrm{g\,cm^{-2}}$ (run 462P) and $\Sigma=191$ $\mathrm{g\,cm^{-2}}$ (run 435). Figures \ref{435F}-\ref{462F} show the temporal evolution of $\tilde{\alpha}$, $\braket{R_{\rm m}}_{x,y}$, and $\braket{B_y}_{x,y}$ for those two runs before and after resistivity is switched on.

Turning on resistivity has no influence on run 435PR (Fig.~\ref{435F}). $R_{\rm m}$ stays well above the critical value of $10^{4}$ in the midplane. Some regions above the photosphere have a lower $R_\mathrm{m}$, without consequences as most of the heating due to MRI happens in the densest part of the disk.

Run 462P has a lower $\Sigma$ and temperature ($[T_{\rm mid}]\approx 3100\rm\,K$ compared to 3750\,K for 435P). Here, $R_{\rm m}$ starts to drop as soon as  resistivity is switched on (Fig.~\ref{462F}). The turbulence is suppressed once $R_\mathrm{m}$ decreases below $\approx 5000$ and   the transport of angular momentum ceases. $\braket{B_y}_{x,y}$ diffuses as expected: first near the photosphere and then, as $R_\mathrm{m}$ decreases, in the whole box. 


\subsection{Influence of ambipolar diffusion and Hall effect}
Let us first compare the order of magnitude of Ohmic (O), ambipolar (A) and the Hall effect (H). From \cite{balbus2001}, we have
\begin{align}
\frac{O}{H}&=\Bigg(\frac{\rho}{3\times 10^{-6}\ \mathrm{g.cm}^{-3}}\Bigg)^{1/2}\Bigg(\frac{c_s}{v_\mathrm{A}}\Bigg),\\
\frac{O}{A}&=\Bigg(\frac{\rho}{10^{-8}\ \mathrm{g.cm}^{-3}}\Bigg)\Bigg(\frac{T}{10^3\ \mathrm{K}}\Bigg)^{-1/2}\Bigg(\frac{c_s}{v_\mathrm{A}}\Bigg)^2.
\end{align}

We take 462PR as a fiducial run (as it is the first one which is altered by Ohmic diffusion) to compare the importance of each non-ideal effect. The typical values deduced from this run for the density, temperature, ionization fraction and magnetisation $v_\mathrm{A}/c_s$ are, respectively, $5\times10^{-7}\:\mathrm{g\,cm^{-3}}$, 3000\,K, $10^{-5.5}$ and $0.1$. Therefore, in this run, ambipolar diffusion is expected to have a negligible impact while the Hall effect is only an order of magnitude weaker than Ohmic diffusion. Note that these conclusions also holds as the disc gets cooler when MRI is suppressed. 

However, comparing the amplitude of each non-ideal effect might not be enough to conclude on the dynamics since they affect the MRI in different ways. Let us therefore analyse individually the impact of ambipolar diffusion and of the Hall effect.

Ambipolar diffusion is characterized by the ambipolar Elsasser number \citep{balbus2001}:
\begin{equation}
\Lambda_\mathrm{A}\equiv\frac{\gamma_i\rho_i}{\Omega}
\end{equation}
where $\rho_i=n_im_i$ is the ion density and $\gamma_i= \braket{\sigma v}_\mathrm{ni}/(m_\mathrm{n}+m_\mathrm{i})=2.7\times10^{13}\big(41.33m_\mathrm{u}/m_\mathrm{n}+m_\mathrm{i}\big)\:\mathrm{cm^3\,g^{-1}\,s^{-1}}$ is the ion-neutral drag coefficient. Following  \citet{balbus2001}, we assume a momentum rate coefficient for ion-neutral collision $\braket{\sigma v}_\mathrm{ni}=1.9\times10^{-9}\:\mathrm{cm^3\,s^{-1}}$ \citep{draine2010}, $m_\mathrm{n}=2.33m_\mathrm{u}$ and $m_\mathrm{i}=39m_\mathrm{u}$ (the neutral and ion masses respectively). We obtain $\Lambda_\mathrm{A}\sim10^3$ for our typical values. \cite{hawley1998} and \cite{bai2011} found that angular momentum transport is not significantly impacted when $\Lambda_{A}>100$. Their result is based on numerical simulations with a net magnetic flux and should still hold when there is no net flux: a net magnetic flux implies currents outside of our domain of simulation that are not affected by resistivity and help sustain MRI turbulence. We conclude the influence of ambipolar diffusion would have on run 462PR is negligible compared to Ohmic diffusion.

The Hall effect is characterized by the Hall Lundquist number 
\begin{equation}
\mathcal{L}_\mathrm{H}\equiv\sqrt{\frac{4\pi}{\rho}}\frac{n_e eH}{c}.
\end{equation}

The impact of the Hall effect on the MRI saturation level is two-fold. For strong Hall effect ($\mathcal{L}_\mathcal{H}\lesssim 5$), turbulent transport is essentially suppressed and the flow relaxes into a self-organised state \citep{kunz2013}. For weaker Hall effect ($\mathcal{L}_\mathcal{H}\gtrsim 10$) the system produces sustained turbulent transport provided that the magnetic reynolds number is sufficiently large, similarly to the pure Ohmic case. This critical reynolds number for sustained turbulence is reduced to $10^3$ for $\mathcal{L}_\mathcal{H}\simeq 30$ while it is identical to the pure Ohmic case for $\mathcal{L}_\mathcal{H}\gtrsim 60$ \citep{sano2002}, indicating that the Hall effect becomes negligible above this threshold.

 Our typical values imply $\mathcal{L}_\mathrm{H}\sim10^{3}$, so the Hall effect does not affect the MRI close to the region where Ohmic diffusion becomes problematic. As the discs cools down in the MRI stable region, $\mathcal{L}_\mathrm{H}$ decreases, making the Hall effect stronger and potentially reviving the MRI. However, the ratio $\mathcal{L}_\mathrm{H}/R_{\rm m}$ does not depend on the temperature, so that at the temperature where $\mathcal{L}_\mathrm{H}\lesssim 50$, one also expects $R_{\rm m}\lesssim 50$, well below the critical $R_{\rm m}$ found in Hall-MRI simulation \citep{sano2002}. We conclude that the Hall effect does not affect the critical temperature (or $\Sigma$) below which DNe are MRI-stable, nor revive the MRI at lower temperatures.\\

\subsection{Influence of X-ray irradiation}
The ionization fraction $x_e$ is critical to determine whether or not the coupling between fluid and magnetic field is strong enough for MRI turbulence to be active. In the resistive runs described above (\S\ref{sec:decay}) $x_e\approx 4\times 10^{-6}$ in the MRI-stable run 462PR (at midplane) and $x_e\approx 6\times 10^{-6}$ in the MRI-unstable run 435PR. Therefore, a very small ionization fraction of order $5\times 10^{-6}$ can be sufficient to maintain turbulence. These ionization fractions are thermal and do not take into account external sources of ionizations. In particular, quiescent DNe show hard X-ray emission originating from the accretion boundary layer onto the white dwarf that may provide sufficient ionization to maintain the MRI active in the quiescent disk. 

The quiescent X-ray spectra are bremsstrahlung with temperatures $kT_{XR}$ ranging from 1 to 10\,keV and luminosities $L_{XR}$ ranging from $10^{28}$ to $10^{32}$\,erg\,s$^{-1}$  \citep{byckling2010}, properties akin to those of a protostar. The typical surface densities under consideration are similar to those of a protostellar disk at 1 AU but the DNe disk is $\sim 10^3$ closer to the X-ray source, so the impinging ionizing flux is much higher. 

We calculate the expected ionization fraction $x_e$ by following studies of protoplanetary disks. We first compute the ionization rate $\zeta$ as in \citet{1997ApJ...480..344G,1997ApJ...485..920G}, assuming a photon flux
\begin{equation}
f_0 = {\cal C}\frac{L_{XR}}{4\pi R_0^2kT_{XR}}
\end{equation}
where ${\cal C}$ is a factor taking into account the irradiation geometry and the albedo $A$ of the disk. The  X-ray emission region is about the size of the white dwarf \citep{mukai2017}, which is much greater than the disk height. In this case, we may expect \citep{1997MNRAS.288L..16K}
\begin{equation}
{\cal C}\approx\frac{2}{3\pi}\left(\frac{R_{WD}}{R_0}\right) (1-A) \approx 10^{-2} (1-A)
\end{equation}
with a white dwarf radius $R_{WD}=10^9$\,cm. 

We obtain $x_e$ by solving Eq.~11 from \citet{fromang2002}, which relates $x_e$ to the  fraction of metals $x_M=n_M/n_n$, the dissociative recombination rate for molecular ions, the radiative recombination rate for metal atoms and the rate of charge exchange between them. We further assume that the disk is passive i.e. has no internal heating and the only source of heat is X-ray irradiation. Hence, in steady-state, the disk is isothermal at the temperature set by $\sigma T^4_\mathrm{irr}=kT_{XR} f_0$. The vertical structure is in hydrostatic equilibrium with $\Sigma=100\,\mathrm{g\,cm^{-2}}$ to compare to our results with the first MRI-stable run, 462PR.

We find that X-ray ionization is negligible, in agreement with the order-of-magnitude estimate of  \citet{Gammie1998}. For example, with ${\cal C} L_{XR}=10^{30}\rm\,erg\,s^{-1}$ and $kT_{XR}=10$\,keV, we find $\zeta=3.4\times10^{-9}\rm\,s^{-1}$ at the disk midplane. The isothermal atmosphere has  $T=1700\rm\,K$ and a midplane density $n=3.7\times 10^{17}\rm\,cm^{-3}$, giving $x_e=4.1\times 10^{-10}$ when no metals are present ($x_M=0$) and   $x_e=1.1\times 10^{-7}$ when $x_M=6.86\times 10^{-6}$ (solar abundance). The fraction $x_e$ decreases for lower values of $\cal C L_{XR}$ or $kT_{XR}$. Hence, X-ray ionization is unable to provide the required ionization fraction  except in the upper layer. However, only a small amount of material is active in this layer: taking the values assumed above, $x_e>5\times 10^{-6}$ for $z>1.5\times 10^8\rm\,cm$ (about 2$H$) and the column density of the active layer represents less than 3\% of $\Sigma$. 

\section{Conclusions}
We have performed stratified, radiative, ideal and resistive MHD, shearing box simulations in conditions appropriate to DNe. We find that the thermal equilibrium solutions found by the simulations trace the well-known S-curve derived from $\alpha$-prescription models, including a middle branch that extends the cold branch to higher temperatures and is characterized by vigorous convection.  We confirm that $\alpha$ increases to $\approx 0.1$ near the unstable tip of the hot branch as reported by \citet{Hirose2014}. This increase is thus robust against the choice of numerical code and, as we investigated, against  the choice of outflow or periodic vertical boundary conditions. Although convection plays a major role in transporting heat in the runs with an enhanced $\alpha$, we find no clear relationship between $\alpha$ and $f_{\rm conv}$, the average fraction of the flux carried by convection. Notably,  $\alpha$ is not enhanced on the middle branch although it is strongly convectively-unstable. Detailed studies of the interplay between convection and the MRI dynamo will be required to understand the exact mechanism behind the enhancement of $\alpha$ towards the tip of the hot branch.

Ohmic dissipation can prevent MRI-driven turbulent transport in the cold, quiescent state of DNe \citep{Gammie1998}. We verified that Ohmic dissipation is the dominant non-ideal effect and carried out resistive MHD simulations to test its influence on $\alpha$.  We find that the region of the cold branch with  $\Sigma<191\,\mathrm{g\,cm^{-1}}$ does not maintain the MRI-driven turbulence.  \citet{Fleming2000} found in isothermal simulations that the presence of a net vertical magnetic flux, providing external support to the MRI, sustains turbulence down to $R_\mathrm{m}=100$. Net flux may help push the MRI-inactive region to slightly lower $\Sigma$ but the dynamical consequences might be difficult to evaluate without a global model to study the evolution of the net flux. A net magnetic field is also expected to impact substantially the saturated value of $\alpha$ when the corresponding $\beta\leq$ is $10^5$ (\citealt{bai2013}; \citealt{salvessen2016}), i.e. $B\geq20\:G$ on the hot branch at a $\Sigma$ of $174\:\mathrm{g.cm^{-2}}$ and $B\geq4\:G$ on the cold branch at a $\Sigma$ of $93\:\mathrm{g.cm^{-2}}$. Using a dipole approximation, this corresponds to a surface magnetic field of $\sim10^{4-5}\:G$ for the white dwarf, smaller than the typical field of intermediate polars (in which the disk is truncated by the white dwarf magnetosphere), or $10^{2-3}\:G$ for the companion, somewhat on the high side of the measured magnetic fields in low-mass stars (see Figure 3 of \cite{donati2009}). Hence, in principle, the binary components might be able to provide enough net magnetic flux to impact transport even though, in practice, how this is achieved is unknown. 

A quiescent disk will not accrete if it enters the MRI-inactive region. This is problematic because the X-ray emission observed in quiescent DNe requires ongoing accretion onto the white dwarf. The X-ray emission region is constrained to the immediate vicinity of the white dwarf by eclipses, as expected from the accretion boundary layer (see \citealt{mukai2017} and references therein). We verified that this emission is unable to self-sustain MRI-driven accretion by photo-ionizing the disk.

If MRI remains inactive, angular momentum transport may be due to an entirely different mechanism in quiescence. Spiral shocks are often invoked as a mean to accrete. They have been proposed to explain the observed anti-correlation between outburst recurrence time and binary mass ratio in some subtypes of dwarf novae (\citealt{cannizzo2012}; \citealt{menou2000}). However, other parameters, such as the mass transfer rate from the secondary, also depend on the mass ratio and therefore impact accretion, rendering the interpretation of this anti-correlation difficult. Besides, global simulations studying the propagation of spiral shocks in hydrodynamics (\citealt{savonije1994}; \citealt{lesur2015}; \citealt{arzamasskiy2017}) and MHD \citep{ju2017} showed that when the ratio $H/R$ is of the order of $10^{-2}$, as is the case in the cold state (especially in the outer region which is the region of interest as spiral shocks propagate inward), spiral shock dissipate rapidly and do not lead to any accretion in the inner parts of the disk. 

The problem may turn out to be a red herring if the disk actually never samples the regime where MRI is inactive. Our results show that accretion is possible on the high $\Sigma$ part of the cold branch, and this may be enough for the DNe cycle to operate. DIM models of DNe outburst cycles with an $\alpha$-prescription show disk rings follow complex tracks in the $(\Sigma, T)$ plane when a cooling front propagates through. The ring does not simply cool on a thermal timescale at constant $\Sigma$ from the tip of the hot branch to the cold branch of the S-curve. Radial heat fluxes cause it to follow a sideways trajectory to a higher $\Sigma$ on the cold branch (see e.g. Fig.~9 in \citealt{1999MNRAS.305...79M}), and this would ensure it never samples the MRI-inactive part of the cold branch. These radial fluxes become important only because the cooling front size is of order of the vertical height of the disk, a situation that strains the assumption in these calculations that the radial and vertical directions are decoupled. Consequently, there are some uncertainties on how they are taken into account by the DIM and their influence on the lightcurves \citep{2010ApJ...725.1393C}; uncertainties that only a global model of DNe outbursts under MRI-driven transport could resolve. 

More prominently, such a global model would also address whether realistic lightcurves of DNe outbursts can be obtained at all with MRI transport. In this regard, the enhancement of $\alpha$ up to the values required by the DIM is both a major advance and a set back. A set back because the high values of $\alpha$ are limited to the tip of the hot branch, whereas DIM models assumed $\alpha$ to be increased over the whole hot branch. The difference has important consequences on the lightcurves predicted by the DIM when $\alpha(T)$ is taken from the MRI simulations: \citet{2016MNRAS.462.3710C} obtained very jagged lightcurves that, in our opinion, compare less favourably to the observations than those of the initial DIM  (see their Fig.~9). Yet, this is also undisputedly a major advance because this at last provides a physical grounding to the phenomenological change in $\alpha$ between hot and cold state first noticed $>$30 years ago; and because, after a certain lull, interest in dwarf novae, the source of much of the basic understanding of accretion physics, has been re-awakened.

\begin{acknowledgement}
We thank the referee for valuable comments and suggestions. NS acknowledges financial support from the pole PAGE of the Universit\'e Grenoble Alpes. GL, GD and MF are grateful to the participants of the KITP 2017 program on {\em Confronting MHD Theories of Accretion Disks with Observations} for the many useful conversations pertaining to this work (and thus this research was supported in part by the National Science Foundation under Grant No. NSF PHY-1125915). This work was granted access to the HPC resources of IDRIS under the allocation A0020402231 made by GENCI (Grand Equipement National de Calcul Intensif). Some of the computations presented in this paper were performed using the Froggy platform of the CIMENT infrastructure (https://ciment.ujf-grenoble.fr), which is supported by the Rhône-Alpes region (GRANT CPER07\_13 CIRA), the OSUG@2020 labex (reference ANR10 LABX56) and the Equip@Meso project (reference ANR-10-EQPX-29-01) of the programme Investissements d'Avenir supervised by the Agence Nationale pour la Recherche.
\end{acknowledgement}

\bibliographystyle{aa}
\bibliography{biblio}

\end{document}